\documentclass{emulateapj}

\usepackage{natbib}
\begin{document}

%
%
\def \dlow {\mbox{$400 {\rm ~l~mm}^{-1}$}}
\def \dhigh {\mbox{$600 {\rm ~l~mm}^{-1}$}}
\newcommand       \be           {\begin{equation}}
\newcommand       \ee           {\end{equation}}
\newcommand       \cm           {\,{\rm cm }}
\newcommand       \pc           {\,{\rm pc }}
\newcommand       \yr           {\,{\rm yr }}
\newcommand       \kpc          {\,{\rm kpc }}
\newcommand       \K            {\,{\rm K }}
\newcommand       \Jy           {\,{\rm Jy }}
\newcommand       \Hz           {\,{\rm Hz }}
\newcommand       \erg          {\,{\rm erg }}
\newcommand       \ergs         {\,{\rm erg \,\, s}^{-1}}
\newcommand       \Myr          {\,{\rm Myr }}
\newcommand       \mgii         {{\rm Mg_{\rm II}}}
\newcommand       \vth          {v_{\rm th}}
\newcommand       \vl           {v_{\rm line}}
\newcommand       \taumg        {\tau_{\rm 2796}}
\newcommand       \sigmg        {\sigma_{\rm 2796}}
\newcommand       \nmgii        {n_{\rm MgII}}
\newcommand       \nmgone       {n_{\rm MgII}^{\tau=1}}
\newcommand       \sigth        {\sigma_{\rm 2796}^{\rm th}}
\newcommand       \lsob         {l_{\rm Sob}}
\newcommand       \leff         {l_{\rm eff}}
\newcommand       \epmg         {\epsilon_{\rm Mg^+}}
\newcommand       \mdotout      {\dot M_{\rm outflow}}
\newcommand       \mcl          {M_{cl}}
\newcommand       \dotms        {{\dot M_*}}
\newcommand       \ls           {l_s}
\def\etal{{\it et al.\thinspace}}
\def\-{{\em{---}}}
\def \mA {\mbox{${\rm m \AA} $} }
\def \rr {\mbox{${\rm RR}$} }
\def \rarb {\mbox{${\rm R_AR_B}$} }
\def \rara {\mbox{${\rm R_AR_A}$} }
\def \dd {\mbox{${\rm DD}$} }
\def \dada {\mbox{${\rm D_AD_A}$} }
\def \dadb {\mbox{${\rm D_AD_B}$} }
\def \dr {\mbox{${\rm DR}$} }
\def \darb {\mbox{${\rm D_AR_B}$} }
\def \dara {\mbox{${\rm D_AR_A}$} }
\def \dbra {\mbox{${\rm D_BR_A}$} }
\def \hMpc      {h^{-1}{\rm\ Mpc}}
\def \hkpc      {h^{-1}{\rm\ kpc}}
\def \h         {\hbox{$\, h$} }
\def \hinv      {\hbox{$\, h^{-1}$} }
\def \hinvseven    {\hbox{$\, h_{70}^{-1}$} }
\def\ewr{\mbox {EW$_r$}}
\def\ewo{\mbox {EW$_o$}}
\def\H7{\mbox {$h_{0.7}$}}
\def\naI{\mbox {\ion{Na}{1}}}
\def\mgI{\mbox {\ion{Mg}{1}}}
\def\feI{\mbox {\ion{Fe}{1}}}
\def\oVI{\mbox {\ion{O}{6}}}
\def\znII{\mbox {\sc Zn~II~}}
\def\crII{\mbox {\sc Cr~II~}}
\def\alI{\mbox {\sc Al~I~}}
\def\alII{\mbox {\sc Al~II~}}
\def\alIII{\mbox {\sc Al~III~}}
\def\mgII{\mbox {\ion{Mg}{2}}}
\def\mnII{\mbox {\ion{Mn}{2}}}
\def\niII{\mbox {\ion{Ni}{2}}}
\def\feII{\mbox {\ion{Fe}{2}}}
\def\feIII{\mbox {\ion{Fe}{3}}}
\def\cIV{\mbox {\ion{C}{4}}}
\def\sV{\mbox {\ion{S}{5}}}
\def\siIV{\mbox {\ion{Si}{4}}}
\def\siII{\mbox {\ion{Si}{2}}}
\def\siI{\mbox {\ion{Si}{1}}}
\def\cII{\mbox {\ion{C}{2}}}
\def\cIII{\mbox {\ion{C}{3}}}
\def\llambda{\mbox {$\lambda$}}
\def\mstar{\mbox {$M_*$}}
\def\hlen{\mbox {$h_{0.7}^{-1}$}}
\def\lstarlya{\mbox {$L^*_{Ly\alpha}$}}
\def\IZw18{I~Zw~18}
\def\m82{M82}
\def\Ab{Abell~}
\def\gi{\mbox {\rm g-i}}
\def\ug{\mbox {\rm u-g}}
\def\br{\mbox {\rm b-r}}
\def\eqn{equation}
\def\vesc{\mbox {$v_{\rm esc}$}}
\def\heha{\mbox {He~I~$\lambda 5876$ / H$\alpha$}}
\def\xhe{\mbox {$\chi({\rm He}) / \chi({\rm H})$} }
\def\heii{\mbox {${\rm He}^+$}}
\def\he{\mbox {\rm He}}
\def\hii{\mbox {${\rm H}^+$}}
\def\h{\mbox {\rm H}}
\def\mab{\mbox {$\rm m_{AB}$}}
\def\ssp{\baselineskip=13pt plus 1pt minus 1pt}
\def\tsp{\baselineskip=5pt plus 1pt minus 1pt}
%
%
\def\deg{\mbox {$^{\circ}$}}
\def\msun{\mbox {${\rm ~M_\odot}$}}
\def\zsun{\mbox {${\rm ~Z_{\odot}}$}}
\def\lsun{\mbox {${~\rm L_\odot}$}}
\def\msunyr{\mbox {$~{\rm M_\odot}$~yr$^{-1}$}}
\def\angs{\mbox {~\AA}}
\def\lya{\mbox {Ly$\alpha$~}}
\def\Ha{\mbox {H$\alpha$~}}
\def\Hb{\mbox {H$\beta$~}}
\def\Hg{\mbox {H$\gamma$~}}
\def\tion{\mbox {$T_{\rm ion}$~}}
\def\ch{\mbox {$\bigtriangleup$}}
\def\grad{\mbox {$\bigtriangledown$}}
\def\lstar{\mbox {$L^*$}}
\def\line{\mbox {~$\lambda$}}
\def\lines{\mbox {~$\lambda\lambda$~}}
\def\h0{\mbox {~H$_0$}}
\def\q0{\mbox {~q$_0$}}
%
%
\def\auroral{[OIII]~$\lambda4363$~}
\def\auroral{[OIII]~$\lambda4363$~}
\def\ohsun{\mbox {(O/H)$_{\odot}$~}}

\def\o3hb{[OIII]$\lambda5007$~/~H$\beta$~}
\def\O1ha{[OI]$\lambda6300$~/~H$\alpha$~}
\def\Ru{[OII]$\lambda\lambda3727$~/~[OIII]$\lambda5007$~}
\def\s2ha{[SII]$\lambda\lambda6717,31$~/~H$\alpha$~}
\def\2z2{HeII~$\lambda4686$~}
\def\z7{[NII]~$\lambda6583$ }
\def\N2{[NII]~$\lambda6583$~/~H$\alpha$~}
\def\16z2{[SII]~$\lambda\lambda6717, 6731$ }
\def\HgI{HgI~$\lambda4358$~}
\def\Sdensity{[SII]~$\lambda6717 / \lambda6731$}
\def\Temp{[OIII]~$\lambda\lambda4959 + 5007 ~{\rm to}~ \lambda4363$~}
%
%
\def\n{NGC~}
\def\asec{\ifmmode {'' }\else $''~$\fi}  
\def\amin{\ifmmode {' }\else $'~$\fi}    
\def\arcsper{\ifmmode \rlap.{'' }\else $\rlap{.}'' $\fi} 
\def\arcmper{\ifmmode \rlap.{' }\else $\rlap{.}' $\fi} 
\def\sles{\lower2pt\hbox{$\buildrel {\scriptstyle <}
   \over {\scriptstyle\sim}$}} 
\def\sgreat{\lower2pt\hbox{$\buildrel {\scriptstyle >}
    \over {\scriptstyle\sim}$}} 
\def\gapp{\mbox {$_>\atop{^\sim}$}}  
\def\lapp{\mbox {$_<\atop{^\sim}$}}  
%
\def\kms{\mbox {~km~s$^{-1}$}}
\def\ergsec{~ergs~s$^{-1}$~}
\def\sb{~ergs~s$^{-1}$~cm$^{-2}$~arcsec$^{-2}$}
\def\flux{~ergs~s$^{-1}$~cm$^{-2}$}
\def\flam{~ergs~s$^{-1}$~cm$^{-2}$ \AA$^{-1}$}
\def\cm3{~cm$^{-3}$}
\def\col{\mbox {~cm$^{-2}$}}
\def\mpc3{~Mpc$^{3}$}
\def\mpc-3{~Mpc$^{-3}$}
\def\rate{~sec$~{-1}$}
\def\um{~${\mu}$m~}
\def\fig{{Figure}}
\def\figs{{Figures}}
\def\tbl{{Table}~}
\def\sec{{Sec.}~}
\def\x{{X-ray}~}
\def\xs{{X-rays}~}
\def\X{{X-Ray}~}

%
\def\et{{\rm et\thinspace al.}\ }   
\def\ets{{\rm et\thinspace al.'s}\ }   
\def\reff{\par\noindent\parskip=1pt\hangindent=3pc\hangafter=1}
\def\apj{ApJ}
\def\apjs{ApJS}
\def\pasp{PASP}
\def\aj{AJ}
\def\mn{MNRAS}
\def\nat{Nature}
\def\aa{A\&A}
\def\aasup{A\&AS}
\def\baas{BAAS}
\def\annrev{ARA\&R}
\def\aar{A\&AR}
\def\pasj{PASJ}
%

%
\def\beginrefs{
         {\normalsize}
         {\noindent}
         \small
        \baselineskip=11pt
        \parindent=0pt
        \frenchspacing
        \parskip=1pt plus 1pt
        \everypar={\hangindent=0.42in}}

\title{Scattered Emission from $z \sim 1$ Galactic Outflows }

\author{Crystal L. Martin\altaffilmark{1} 
  Alice E. Shapley\altaffilmark{2,3}
  Alison L. Coil\altaffilmark{4,5}
  Katherine A. Kornei\altaffilmark{2}
  Norman Murray\altaffilmark{6}
  Anna Pancoast\altaffilmark{1} 
}

\altaffiltext{1}{Department of Physics, University of California, 
Santa Barbara, CA, 93106, cmartin@physics.ucsb.edu}

\altaffiltext{2}{Department of Physics and Astronomy, University of California, 
Los Angeles, CA, 90025}

\altaffiltext{3}{Packard Fellow}

\altaffiltext{4}{Center for Astrophysics and Space Sciences,
Department of Physics, University of California, San Diego, CA 92093}

\altaffiltext{5}{Alfred P. Sloan Fellow}

\altaffiltext{6}{Canadian Institute for Theoretical Astrophysics, 60 St. George Street, University of Toronto, Toronto, ON M5S 3H8, Canada}

\begin{abstract}
Mapping \mgII\ resonance emission scattered by galactic winds offers
a means to determine the spatial extent and density of the warm outflow.
Using Keck/LRIS spectroscopy, we have resolved
scattered \mgII\  emission to the east of 32016857, a star-forming galaxy at $z =0.9392$
with an outflow. The \mgII\ emission from this galaxy exhibits a P-Cygni profile, extends
further than both the continuum and [\ion{O}{2}]  emission along the eastern side of the
slit, and has a constant Doppler shift along the slit which does not follow the velocity
gradient of the nebular [\ion{O}{2}] emission. Using the Sobolev approximation, we
derive the density of ${\rm Mg}^+$ ions at a radius of 12 - 18~kpc in the outflow.
We model the ionization correction and find that much of the outflowing Mg is in ${\rm Mg}^{++}$.
We estimate that the total mass flux could be as large as 330 - 500 \msunyr, with the largest
uncertainties coming from the depletion of Mg onto grains and the clumpiness of the
warm outflow. We show that confining the warm clouds with a hot wind reduces the
estimated mass flux of the warm outflow and indicates amass-loading factor near unity
in the warm phase alone. Based on the high blue luminosities that
distinguish 32016857 and TKRS~4389, described by \cite{Rubin:2011p1660}, from
other galaxies with P-Cygni emission, we suggest that, as sensitivity to
diffuse emission improves, scattering halos may prove to be a generic property of
star-forming galaxies at intermediate redshifts.
\end{abstract}

\section{Introduction}

In the baryon-driven picture of galaxy evolution, gas accretion onto galaxies, 
gas consumption by star formation, and gas ejection by galactic winds shape 
the properties of galaxies. The only simple aspect of this baryon cycle may be
the rate of gas infall into galactic halos, which may follow the mean growth 
rate of the dark matter. How much of the inflow can cool and fuel star formation, 
versus being shock heated to the halo virial temperature, remains a largely 
theoretical debate \citep{Keres:2005p2,Dekel:2006p2,Keres:2009p160,
Dekel:2009p1612, Bouche:2010p556, vandeVoort:2011p2458, Dave:2012p98}. 
In practice, it has generally been more straightforward to detect outflowing gas
than inflowing gas.

Over a broad redshift range, spectroscopic surveys have empirically described the 
galaxy population hosting outflows using blueshifted, resonance absorption lines, where 
the sign of the Doppler shift (relative to the galactic barycenter) unambiguously identifies 
outflowing gas on the near side of a galaxy.  
\citep[Rubin \et, in prep]{Shapley:2003,Martin:2005,Tremonti:2007p77,Sato:2009p214,
Weiner:2009p187,Chen:2010p445,Steidel:2010p420,Heckman:2011p1481,Coil:2011p46,
Erb:2012p26,Martin:2012}. 
These data provide an accurate picture of which galaxies host outflows,
e.g., namely those with large concentrations of massive stars \citep{Heckman:2003p47,
Kornei:2012p135,Law:2012p29} consistent with theoretical arguments requiring
massive star clusters to generate outflows \citep{Murray:2011p66}. 

The properties of these outflows, in contrast to their demographics, remain poorly 
constrained. The outflowing mass flux, for example, is central to galaxy evolution 
models; yet theoretical models that fit the empirically determined mass function and 
mass -- metallicity relation for galaxies require a mass-loss rate of order the star 
formation rate (SFR) at mass scales comparable to the Milky Way
\citep{Dave:2012p98,Shen:2012p50,Creasey:2013p437,Hopkins:1301.0841,Puchwein:2013p2966}.
The location, maximum velocity, and total column density of the warm outflowing gas 
are not well-constrained by observations of optical and ultraviolet resonance lines in 
galaxy spectra. The absorbing clouds may lie anywhere along the sightline to the galaxy,
and the decline in gas covering fraction with increasing blueshift makes it challenging 
to detect the highest velocity gas \citep{Martin:2009}. 
The hotter wind fluid, which entrains this warm, low-ionization gas,
largely eludes direct detection at intermediate redshifts \citep{Tripp:2011p952}.

Two complementary strategies have emerged recently for measuring the spatial extent of 
low-ionization outflowing gas. First, at large radii, outflows can 
be mapped with sightlines to background quasars or galaxies. Halo absorption from 
outflows may be distinguished from gas accretion by its location in bipolar flows
which emerge roughly perpendicular to the galactic disk
\citep{Bordoloi:2011p1640,Bouche:2012p801,Kacprzak:2012p7}.  
Second, although the emission measure of low density gas
typically makes halo gas undetectable by direct imaging, photons scattered by halo
gas can be directly imaged \citep{Steidel:2011p160,Rubin:2011p1660,Prochaska:2011p24}. 
Combining the information contained in galaxy absorption spectra, resonance emission maps,
and sightlines probing galaxy halos should  greatly improve the accuracy of the
estimated mass-loss rates in outflows.

The most familiar example of resonance scattering involves the \lya photons produced 
in \ion{H}{2} regions, which are scattered in our direction by interstellar and 
circumgalactic gas. P-Cygni \lya line profiles provide direct kinematic evidence 
for scattering by outflowing gas. In the simplest model, the redshifted emission component 
is scattered off the back side of an expanding shell while the near side imprints 
blueshifted, resonance absorption \citep{Pettini:2002p742,Verhamme:2008p89}. Since 
spectra of the bluer, less-obscured high-redshift galaxies \citep{Kornei:2010} show
these P-Cygni line profiles more frequently than do spectra of more obscured galaxies, 
the detection of diffuse \lya emission in stacks of
all types of high-redshift, star-forming galaxies was remarkable \citep{Steidel:2011p160}. 
These halos are apparently common, extend beyond the UV-continuum emission, and indicate
that circumgalactic gas scatters \lya photons which have escaped from the interstellar
medium.

We suggest a physical analogy exists between 
the prominent \mgII\ $\lambda \lambda 2796.35, 2803.53$ P-Cygni line profiles recently
discovered in near-UV spectra of star-forming galaxies \citep{Tremonti:2007p77,
Martin:2009,Weiner:2009p187,Prochaska:2011p24,Rubin:2011p1660,Erb:2012p26} and
the \lya halos detected around higher redshift galaxies. 
The \mgII\ emission is  prominent in bluer galaxies and galaxies with relatively 
low stellar mass \citep{Martin:2012}, although further analysis 
indicates the strongest correlation with specific SFR \citep{Kornei:2013}.
These results appear consistent with the ideas that dust attenuation halts the escape of
\mgII\ photons and outflows play a significant role in shaping the often-observed P-Cygni
profiles.

Motivated by the need to improve empirical mass-loss rates and the discovery of \mgII\ emission 
around TKRS 4389 \citep[z=0.69]{Rubin:2011p1660}, we examined the spatial extent of \mgII\ emission 
in LRIS spectra of individual, blue galaxies at $0.4 < z < 1.4$ included in the outflow census of 
\cite{Martin:2012} and originally discovered by the Deep Extragalactic Evolutionary Probe 2 survey, 
DEEP2 \citep{Newman:2012}.  The spectrum of the outflow galaxy 32016857 resolves extended \mgII\ 
emission which we argue most likely reflects stellar continuum and \ion{H}{2} region emission
scattered by the outflow.  We also describe extended \mgII\ emission around 32010773 and 12019973,
two galaxies with spectra indicating the presence of outflowing gas (blueshifted resonance absorption 
and P-Cygni \mgII\ emission);  but, in contrast to 32016857, the \mgII\ surface brightness profiles
and kinematics can be fitted with either resonance scattering or direct emission from \ion{H}{2}
regions. Finally, the spectrum of 22028686 resolves resonance absorption across the 
galaxy, yet we detect the resonance emission over just one side of the galaxy, a 
situation which may provide insight into the conditions required for resonance emission to
escape.

We use resonance scattering of the continuum light into our sightline
to estimate mass-loss rates for 32016857 and TKRS~4389. To outline the physical 
argument, we consider spherical outflow geometries subtending a solid angle $\Omega$. 
Continuum photons emitted at increasingly higher frequencies (relative to  the resonance) 
are absorbed by atoms at progressively larger radii if the radial velocity, $v(r)$, of the 
outflow increases outwards. At any radius, the spatial extent of 
the scattering region (over a fixed frequency range) grows with increases in the thermal or 
turbulent velocity. Provided the outflow density and the velocity component along 
the sightline do not change much over this interaction region, the computation of 
the scattering optical depth can be simplified using the Sobolev approximation 
\citep{Sobolev:1960,Prochaska:2011p24}. Measurements of the extent of the scattering halo
and the outflow velocity therefore effectively constrain
the ionic density of the outflowing gas at a radius where the scattering 
optical depth is close to unity. We estimate the ionization correction by calculating
the photoionization equilibrium and demonstrate how the clumpiness of the outflow
affects the estimated mass-loss rate.

The paper is organized as follows.
We review the data in Section~\ref{sec:data} and present results for
32016857 in Section~\ref{sec:32016857}. In Section~\ref{sec:halos_general}, we examine
what galaxy properties favor scattered halo gas and present other examples
of spatially extended emission-line gas from our study. We return to 32016857
in Section~\ref{sec:discussion}, where we use the Sobolev approximation 
to estimate the ionic gas density of the outflow, present the photoionization models,
and discuss the implications for the mass-loss rate. An Appendix illustrates how 
the clumpiness of the warm phase of the outflow may affect the estimated mass-loss
rate. 

We adopt a 
Hubble constant of $H_0 = 70$\kms ~Mpc$^{-1}$, and we assume a $\Lambda$CDM 
cosmology with $\Omega_M = 0.3$ and $\Omega_{\Lambda} = 0.7$. This cosmology
yields a scale of 7.884 $h_{70}^{-1}$~kpc/\arcsec\ at redshift $z = 0.9392$.
We use vacuum wavelengths throughout to refer to both ultraviolet and optical
transitions and atomic data from \cite{Morton:2003p498}. Stellar masses
and SFR's are derived assuming a Chabrier initial mass function \citep{Chabrier:2003p763}.

\section{Data} \label{sec:data}

We previously described 208 Keck LRIS spectra of $0.4 < z < 1.4$ galaxies in \cite{Martin:2012}.
Among the 145 spectra with both \mgII\ coverage and fitted \feII\ resonance absorption, 
we have objectively searched for \mgII\ emission \citep{Kornei:2013}. We use
the parameters fitted to the \feII\ absorption troughs in \cite{Martin:2012} to describe
the shape of the intrinsic \mgII\ absorption trough and definitively detect \mgII\ $\lambda 
\lambda 2796, 2803$ resonance emission in 22 spectra. A higher fraction of the spectra may 
have \mgII\ emission which is not uniquely separated from the absorption troughs. 

The primary spectrum discussed in this paper, 32016857, was obtained on mask msc32aa
(see Table 1 in \citep{Martin:2012})
using the configuration with the D560 dichroic, 600~l~mm$^{-1}$ blue grism blazed at
4000 \AA, and the 600~l~mm$^{-1}$ red grating blazed at 7500 \AA. Due to the non-photometric 
observing conditions for this mask, absolute flux calibrations for the blue and red spectra were obtained, 
respectively, by matching the DEEP2 photometry in the $B$ and $I$ bands, a procedure which 
also corrects the broad band luminosity for slit losses. This absolute flux calibration
is less accurate than that derived from observations of standard stars under photometric 
conditions for most masks; however, all the LRIS spectra have accurate 
relative fluxes independent of observing conditions. We refer the interested reader to
Section 2.2 of \cite{Martin:2012} for a more detailed description of the instrumental
configuration, data acquisition, and spectral reduction.

To search for extended emission in the 2D spectral images, 
we fit a linear continuum through bandpasses to either side of the \mgII\ doublet. 
We visually inspected the 2D spectra both before and after continuum 
subtraction for any sign of \mgII\ emission. We collapsed the line emission (in both 
transitions) in the dispersion direction to create a  surface brightness profile and 
compared it to the instrumental point-spread function (defined by a stellar spectrum). 
A similar procedure was followed to create an [\ion{O}{2}] surface brightness profile
from the red spectra. Finally, profiles of the continuum surface brightness were constructed
from both the blue and red spectra directly adjacent to the emission line of interest.

\section{Case Study of the  $z = 0.9392$ Galaxy 32016857} \label{sec:32016857}

\begin{figure*}[t]
 \hbox{\hfill \includegraphics[height=18cm,angle=270,trim= 0 25 75 0]{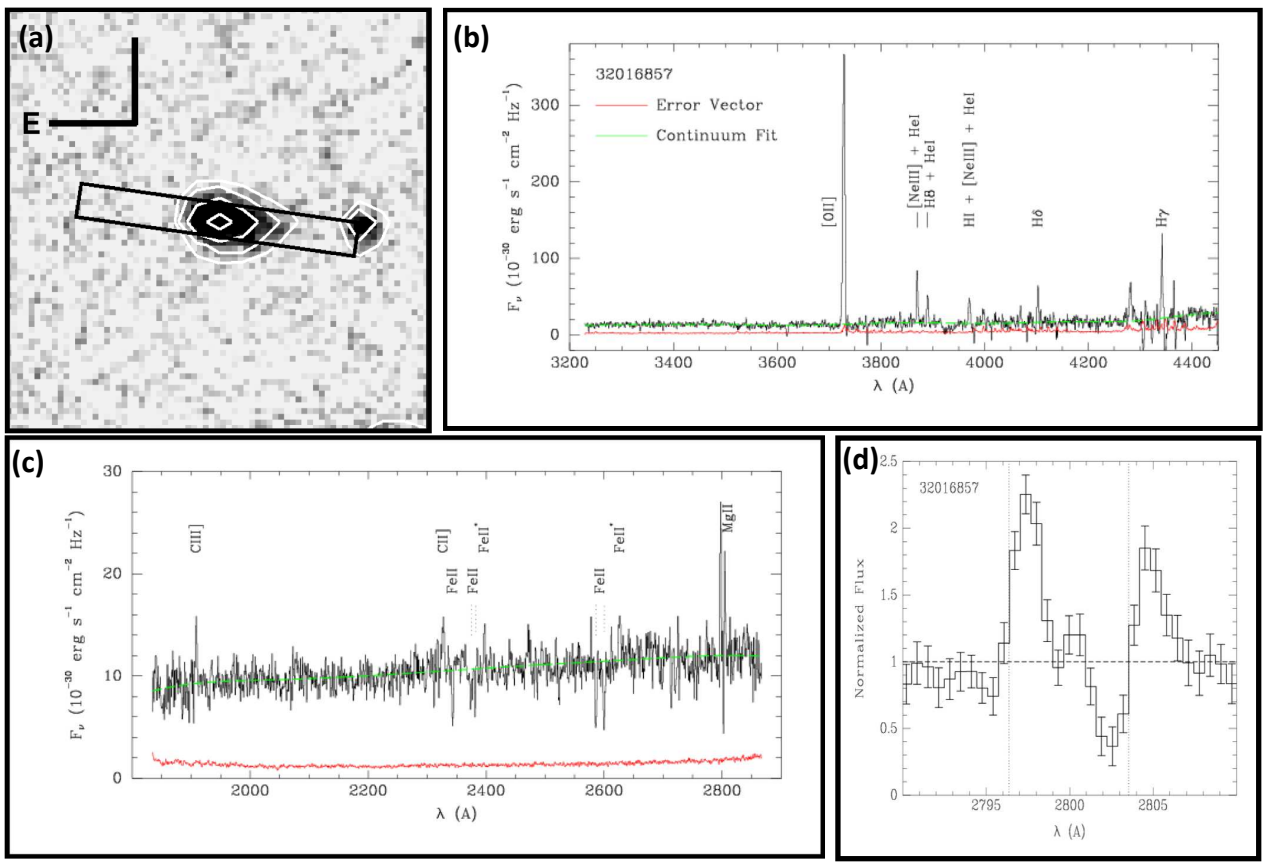}
                \hfill}
          \caption{\footnotesize 
            {\it (a):} CFHT
             $R$-band image of 32016857 from \cite{Coil:2004p765}.
            Contours, spaced by a factor of
            two in surface brightness, show an elongated structure; and 
            the  slit ($PA = 82.0$\deg; 10\farcs0 long by 1\farcs2 wide) runs roughly 
            along the major axis of the galaxy. The surface brightness profile 
            along the major axis, however, is clearly more extended
            to the west than to the east. Due to this asymmetry, an inclined
            disk does not describe the morphology well, and the position
            angle of the semi-minor axis need not indicate the direction of
            the projected rotation axis.
            {\it (b):}
             Red spectrum of 32016857 showing strong lines and fitted continuum level.
            The strong nebular line emission determines the systemic velocity, $z = 0.939196$.
            Table~\ref{tab:lines} provides the transition wavelengths.
           {\it (c):}
            Blue spectrum of 32016857 showing strong absorption lines from \feII,
            \ion{C}{2}] emission, \mgII\ emission, \feII$^*$ emission, and fitted 
            continuum level. The continuum S/N ratio at 2450 \AA\ is 8.3 per pixel.
            Table~\ref{tab:lines} provides the transition wavelengths.
           {\it (d):}
            Integrated spectrum of 32016857 zoomed in to illustrate the P-Cygni-like
            line profile of the \mgII\ doublet. Dotted, vertical lines denote the 
            systemic velocity, and the dashed line marks the normalized continuum level.
}
 \label{fig:intro} \end{figure*}

In this section, we focus on the clearest example of spatially extended, 
scattered \mgII\ emission in our sample, 32016857.  Figure~\ref{fig:intro} 
shows an image of this galaxy as well as the blue and red spectra obtained
previously with LRIS. The $R$-band isophotes of 32016857 indicate a 
major axis roughly aligned with the position angle of the slit. However,
the isophotes clearly show an asymmetry along the major axis which is not consistent 
with the projection of an inclined, symmetric disk. Higher resolution imaging is required 
to distinguish whether the major axis reflects the position angle of a disk or some more 
complicated morphology, such as two merging galaxies. We wish to emphasize, however, that the 
surface brightness distribution and kinematics of the [OII] emission are compatible with two 
emitting regions. In addition, we identify the object 5\arcsec\
west of 32016857 in Figure~\ref{fig:intro}a as 32016527 in the DEEP2 photometric
catalog, which suggests the object is a compact galaxy based on its magnitude and color.

\subsection{Galaxy Properties}

Table~\ref{tab:galaxy} lists the measured properties of 32016857. It
has notably blue $U-B$ color. The SFR estimated from  the [\ion{O}{2}] 
luminosity, as described in the notes to Table 1, is approximately 80\msunyr. 
The  mass derived from the SED fit, $\log (M / \msun) = 9.82$
\citep{Bundy:2006p1663}, places 32016857 in the lowest tertile of the
LRIS sample by stellar mass. This mass estimate should be accurate to within 
a factor of two because the $K$-band photometry of 32016857 directly constrains
the rest-frame, near-IR continuum. At $z \approx 1$, abundance matching places
galaxies with stellar mass $\log M_*/\msun = 9.8$ in halos of total mass
$\log M_h / \msun \approx 11.5$ \citep{Behroozi:2010}, consistent with the clustering of
DEEP2 galaxies which places galaxies brighter than $M_B < -20.77$ in halos more massive
than $\log M_h/\msun = 11.3$ \citep{Coil:2008}. In numerical simulations,
halos with masses similar to our best estimate for 32016857 (specifically
$M_h < 10^{12}$\msun), gas accretion is dominated by the cold flows never 
shock-heated to the halo virial temperature \citep[e.g.,][]{Keres:2005p2}.

In the red side spectrum, shown in panel~b of Figure~\ref{fig:intro}, 
the resolution of 220\kms\ FWHM does not quite resolve the
prominent [\ion{O}{2}] $\lambda \lambda 3727.09, 3729.88$ emission lines. The
blue spectrum, 282 \kms\ FWHM resolution,  reveals strong emission in the \mgII\ 
$\lambda \lambda 2796.35, 2803.53$ doublet with a P-Cygni profile (bottom panels). 
The recombination lines from hydrogen and helium as well as the
strong emission in the \ion{C}{2}] $\lambda 2324-29$, \ion{C}{3}] $\lambda 1908.73$, 
[\ion{Ne}{3}] $\lambda 3869.85, \lambda 3968.58$, and [\ion{O}{2}] lines
come from \ion{H}{2} regions photoionized by massive stars. The absence
of [\ion{Ne}{5}] $\lambda 3426.98$ emission argues against an AGN as the dominant 
source of ionizing radiation.

The emission lines in the red spectrum
dominate the cross-correlation signal with template spectra and thereby determine
the LRIS redshift of $z = 0.9392$. The root-mean-square (RMS) error in the dispersion 
solution of the red spectrum is only a few \kms; hence the  RMS error in the blue
dispersion solution, about 18\kms, determines the systematic uncertainty in the
relative velocities of the low-ionization absorption and nebular emission. 
In the blue spectrum, the centroid of the \ion{C}{3}] emission line lies just
$-30 \pm 31$\kms\ from the intercombination transition which dominates the emission
at high density. We readily
acknowledge a larger discrepancy of 50\kms\ between the LRIS redshift and the DEEP2
survey redshift, likely caused by differences in slit position angle but irrelevant
to the value of relative velocities between the absorption lines and nebular emission.

\begin{figure}[t]
 \hbox{\hfill \includegraphics[height=8cm,angle=270,trim=0 60 0 0]{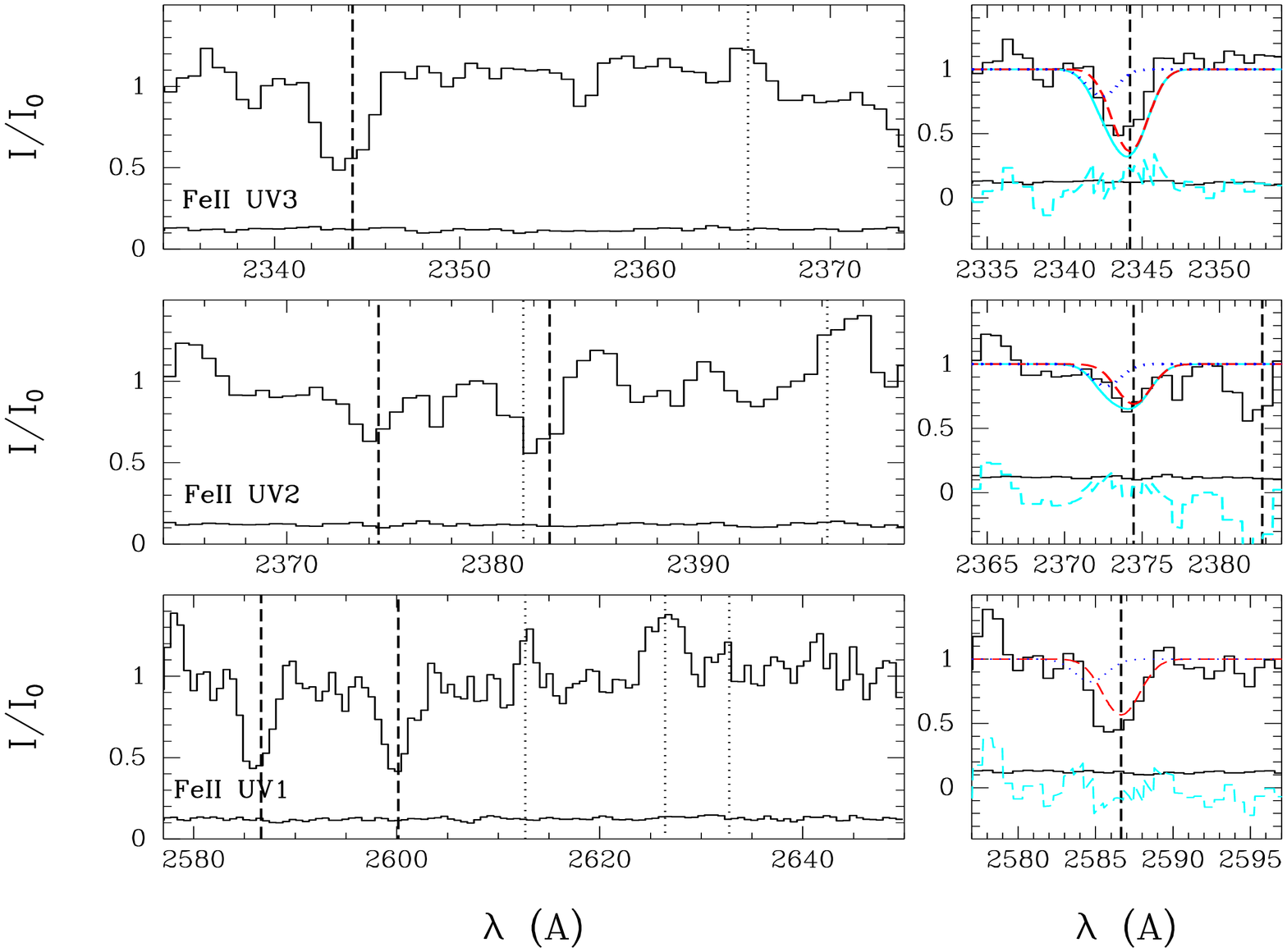}
                \hfill}
          \caption{\footnotesize
            Continuum normalized spectrum showing the near-UV, \feII\ multiplets UV1,
            UV2, and UV3. 
            {\it Left column:}
            The vertical dashed (dotted) lines mark the \feII\ (\feII$^*$) transitions
            listed in Table~\ref{tab:lines}. The $3\sigma$ upper limit on the
            equivalent width of the \feII\ 2261 line, not shown, is 0.60\AA\ 
            \citep{Martin:2012}. Fluorescent emission from the upper level of the
            $\lambda 2600$ transition ($z^6D^0_{9/2}$) is detected at $\lambda 2626$, and
            the marginal detections of both \feII$^* \lambda 2612$ and $\lambda 2632$
            mark fluorescent decays from the $z^6F^0_{7/2}$ level, which is excited
            by absorption in $\lambda 2587$. A net excess of emission is also detected
            near the wavelength of the \feII$^* \lambda 2396$ transition in UV2 and 
            at \feII$^* \lambda 2365$ in UV3. While Table~\ref{tab:lines} shows that
            many of these detections are not individually robust statistically, the
            combination of features clearly argues for strong fluorescent emission 
            in the spectrum of 32016857.
            {\it Right column:}
            Two-component joint fit (solid) to the \feII\ $\lambda$ 2344, 2374, and 2587 transitions.
            The Doppler component (blue, dotted line) has a blueshift of $V_{Dop} = -227 \pm 82$\kms\ and
            an equivalent width  $W_{Dop}(\lambda 2374) = 0.41$\AA. Fit redsiduals (dashed) are shown
            relative to the error spectrum at the bottom of each plot.
             }
 \label{fig:vc_fit} \end{figure}

The absorption lines in the LRIS spectrum are not sufficiently resolved to measure
the maximum blueshift from the \feII\ line profiles. 
The fitted centroid of the \feII\ absorption lines is 
blueshifted $V_1 = -91 \pm 15$\kms \citep{Martin:2012}. The projected outflow
velocity is more accurately estimated by the two-component fit shown in the right column of 
Figure~\ref{fig:vc_fit}. Correcting for the interstellar absorption at the systemic velocity
increases the blueshift of the Doppler component to $V_{Dop} = -227 \pm 82$\kms. 

While the $\lambda 2803$ absorption trough may be filled in by emission from both 
\mgII\ transitions, only $\lambda 2796$ emission fills in the intrinsic $\lambda 2796$
absorption trough. In \mgII\ $\lambda 2803$,  \fig~\ref{fig:intro} shows absorption blueward
of the systemic velocity and redshifted emission. The marginally-resolved emission 
wing extends to +310\kms\ \citep{Martin:2012}.  The corresponding red wing of the 
\mgII\ $\lambda 2796$ profile
may significantly fill in the bluest portion of the intrinsic $\lambda 2803$ absorption 
profile. Emission at the systemic velocity in both transitions likely fills their intrinsic
absorption profiles at the systemic velocity. 

Optically thick \mgII\ absorption will produce an intrinsic $\lambda 2803$ trough
with the same equivalent width as the $\lambda 2796$ trough which is twice as
strong in the optically thin limit. Remarkably, however, the absorption 
equivalent width of the \mgII\ $\lambda 2796$ absorption trough is less than that 
in $\lambda 2803$ as seen in both \fig~\ref{fig:intro} and Table~\ref{tab:lines}.
While low S/N ratio could produce such an effect through measurement error, our
analysis suggests a different interpretation.
We can understand the net \mgII\ profile provided the
intrinsic ratio of emission-line strengths $W_{em}(\lambda2796)/W_{em}(\lambda2803)$ 
exceed the ratio of absorption-line strengths $W_{abs}(\lambda2796)/W_{abs}(\lambda2803)$. 
In support of this approach, we note that in Table~\ref{tab:lines} the net emission 
equivalent width of the stronger \mgII\ transition, $-2.13 \pm 0.29$ \AA, is higher than in 
the weaker \mgII\ line at $\lambda 2803$. Qualitatively at least, we can therefore create a 
line profile with a weak $\lambda 2796$ trough (relative to $\lambda 2803$) if the 
intrinsic absorption troughs are both saturated and therefore equal in area.

\begin{figure}[t]
 \hbox{\hfill \includegraphics[height=6cm,angle=0,trim=0 0 0 0]{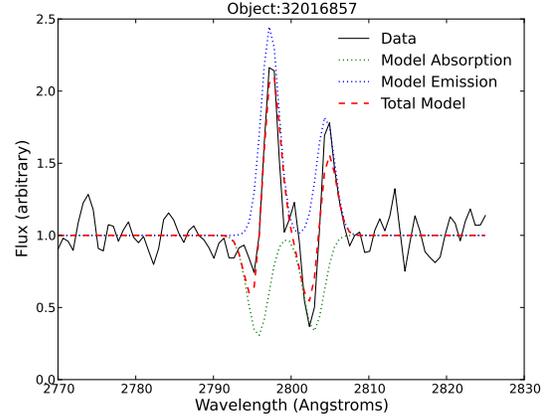}                  \hfill}
\caption{\footnotesize
Typical example of a fit to the \mgII\ complex in the 32016857 spectrum. We model the emission with
two Gaussian line profiles sharing a common Doppler shift and velocity width but independent
amplitudes. We add the emission component to the fitted absorption doublet described by $I(\lambda) = 1 - C_f + C_f
\exp (-[\tau_{2796}(\lambda) + \tau_{2803}(\lambda)])$, where the optical depth in the bluer
line is twice that of the redder line at the same Doppler shift; and a Gaussian distribution describes 
the variation in optical depth with Doppler shift, $\tau(\lambda) = \tau_0 \exp (-[(\lambda - \lambda_0) / 
(\lambda_0 b /c)]^2)$. We convolve this joint model with a Gaussian function describing the instrumental resolution.
The example shown here fit the absorption troughs with a Doppler shift of $V_{abs} = -104$\kms, 
Doppler parameter $b_{Dop} = 70^{+23}_{-24}$\kms, covering factor $C_f = 0.97$, and optical depth at line center $\tau_0 = 15$;
it describes the emission lines with a Doppler shift of $V_{abs} = 70$\kms, width $\sigma = 79$\kms,
amplitude $A_B = 1.67$ for the $\lambda 2796$ transition, and doublet ratio $A_B / A_R = 1.67$. The 
doublet ratio fitted to the emission component is larger than that of the absorption component.
}
 \label{fig:mcmc} \end{figure}

To explore this solution quantitatively in a manner that fully incorporates the
spectral S/N ratio, we fit a model consisting of an emission doublet simply added
to the intrinsic absorption troughs. This model assumes the emission region lies
beyond most of the absorbing gas. The caption of Figure~\ref{fig:mcmc} describes the 
functional form of these components. We wrote a custom Markov Chain Monte Carlo
fitting code in Python to explore the range of possible parameter values. The
fit shown in Figure~\ref{fig:mcmc} illustrates a typical fit drawn from the Markov chain.

To estimate the values of the absorption and emission components, 
we used the posterior probability distributions compiled for each parameter from the full Markov chain 
to determine the mean values of the parameters and their $1\sigma$ uncertainties. 
Our fit describes the absorption troughs with a Doppler shift of $V_{abs} = -11^{+51}_{-49}$\kms,
Doppler parameter $b_{Dop} = 81^{+23}_{-24}$\kms, covering factor $C_f = 0.91^{+0.06}_{-0.12}$, and
optical depth at line center $\tau_0 = 68^{+80}_{-44}$; it finds an emission component with
a redshift of $V_{em} = 105^{+12}_{-10}$\kms, width $\sigma = 14^{+14}_{-9}$\kms,
amplitude $A_{2796} = 4.74^{+0.77}_{-0.81}$ for the $\lambda 2796$ transition, and
doublet ratio $A_{2796} / A_{2803} = 1.51^{+0.20}_{-0.16}$. 
We find that the intrinsic absorption troughs typically have very similar 
equivalent widths. In contrast, only doublet ratios $A_{2796} / A_{2803}$, near 1.5 provide 
acceptable descriptions of the emission. We conclude that the optical depth of the emission region in our
slit is lower than that of the absorbing gas along our sightline. We interpret the difference between 
the absorption and emission doublet ratios as an indication that different physical regions of the 
galaxy produce the emission and absorption -- e.g., non-spherical geometries for the scattering 
region and/or significant \mgII\ emission from \ion{H}{2} regions.

The net \feII\ absorption troughs more nearly reflect the intrinsic shape of the
absorption profile than does \mgII\ absorption. The resonance emission that fills in these
troughs may be generated by \ion{H}{2} regions or created by the absorption of continuum 
photons in the surrounding gas. As emphasized in \cite{Erb:2012p26},
\ion{H}{2} regions produce much more emission in \mgII\ than in \feII, so 
nebular emission will have a proportionately larger impact on the shape of the
net \mgII\ line profile. In our integrated spectra of 32016857, the flux ratio of the 
[\ion{O}{2}] and \mgII\ emission, $F([$ \ion{O}{2} $]~\lambda \lambda 3726,29 ) /
F(\mgII\ \lambda 2796) \approx 42$ (uncorrected for reddening), is consistent with
line ratios calculated for nebulae at slightly sub-solar metallicity photoionized by
a starburst spectral energy distribution. For example, comparison of this line ratio to 
Figure~16 in \cite{Erb:2012p26},  indicates an ionization parameter $\log U \sgreat -2.9$, 
where the lower limit indicates that the reddening-corrected line ratio would require
a larger ionization parameter. In addition, in contrast to the ground state of \mgII, which
is a singlet, the ground-state of \feII\ has fine structure. In \feII, absorption of
a resonance photon is often followed by the emission of a longer wavelength photon
\citep{Rubin:2011p1660,Prochaska:2011p24,Martin:2012}.  This fluorescent emission, clearly 
visible in \fig~\ref{fig:vc_fit} near the systemic velocity, does not fill in the 
resonance absorption troughs. 

Table~\ref{tab:lines} summarizes the equivalent widths and velocities of \feII\ 
absorption troughs and \feII$^*$ emission. The relative strengths of the individual
absorption lines are not well fit with standard curve-of-growth techniques due to
differential amounts of emission filling. Using the two-component fit shown in
\fig~\ref{fig:vc_fit}, \cite{Martin:2012} estimate the ionic column density of
the Doppler component lies in the range  $\log [N_{Dop}(Fe^+) C_f ({\rm ~cm}^{-2})] = 14.42 - 15.53$,
where $C_f$ is a gas covering factor of order unity.

\subsection{Emission Line Properties along the Slit}

\begin{figure*}[t]
 \hbox{\hfill \includegraphics[height=18cm,angle=270,trim=0 0 30 0]{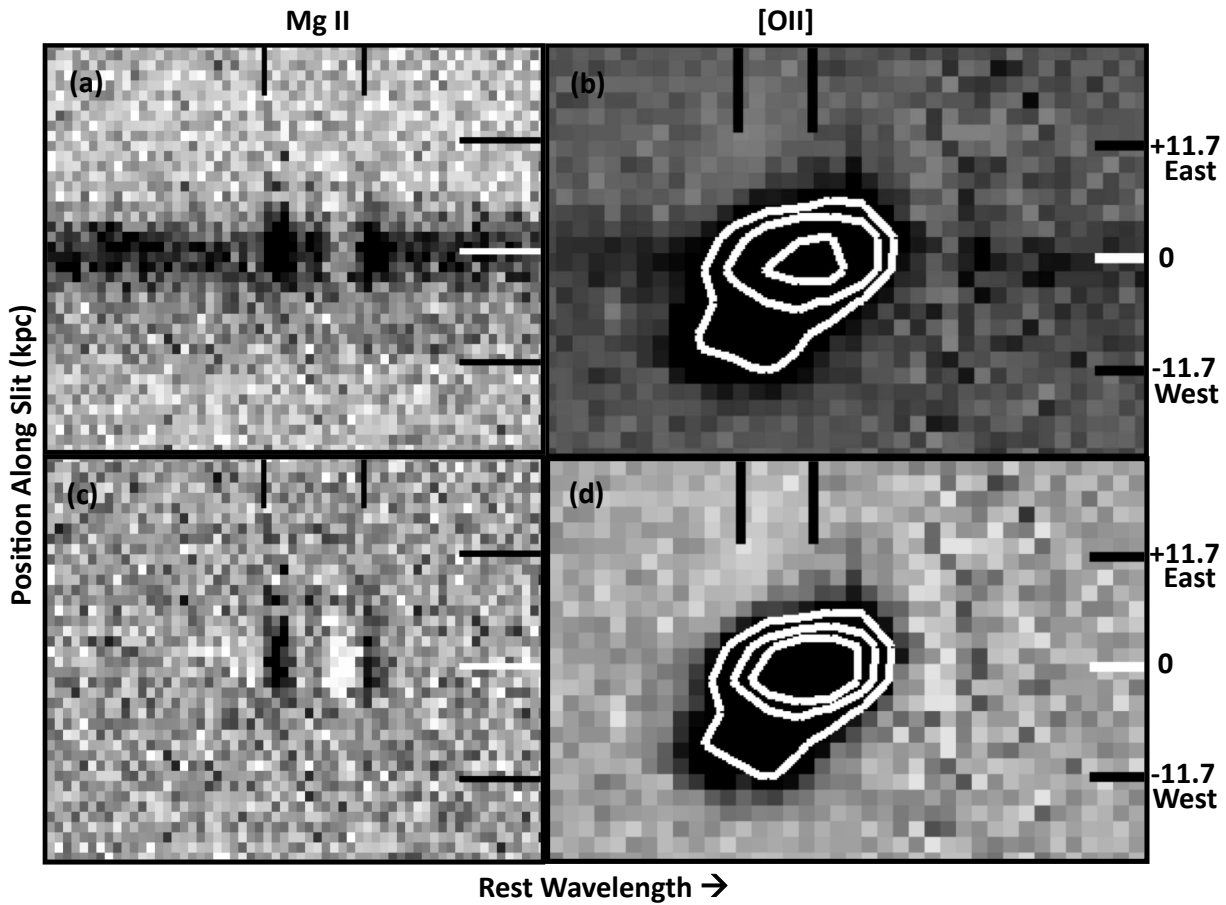}
                \hfill}
          \caption{\footnotesize
            The 32016857 spectrum near \mgII\ and [\ion{O}{2}].
            {\it (a)} The \mgII\ $\lambda\lambda$2796.35, 2803.53  doublet (vertical lines
            separated by 770~km~s$^{-1}$) is detected in emission and absorption.
            {\it (b)} The [\ion{O}{2}] $\lambda\lambda$ 3727.09, 3729.88
            doublet (vertical lines separated by 225\kms) is detected in emission.
            {\it (c)} Continuum subtracted \mgII\ spectrum.
            {\it (d)} Continuum subtracted [\ion{O}{2}] spectrum.
            Linearly spaced contours show a velocity gradient along the slit which we attribute
            to the rotation of the galaxy.
             }
 \label{fig:mosaic} \end{figure*}

The two dimensional (2D) spectra of 32016857 constrain the spatial extent of the emission regions. 
Figure~\ref{fig:mosaic}  compares the structure of the \mgII\ and [\ion{O}{2}]
line emission. While the Doppler shift of the [\ion{O}{2}] emission shifts by
approximately 190 \kms\ over 17 kpc along the slit,  no velocity gradient is
detected in \mgII\ emission. While the \ion{H}{2} regions producing the [\ion{O}{2}]
emission probably contribute to the total \mgII\ emission, based on the kinematic differences
between the 2D spectra, we suggest that the \mgII\ emission takes on the velocity of the 
outflowing material, not the velocity of the \ion{H}{2} regions, because it 
is scattered by halo gas. 

\begin{figure}[t]
 \hbox{\hfill \includegraphics[height=8cm,angle=-90,trim=0 0 0 0]{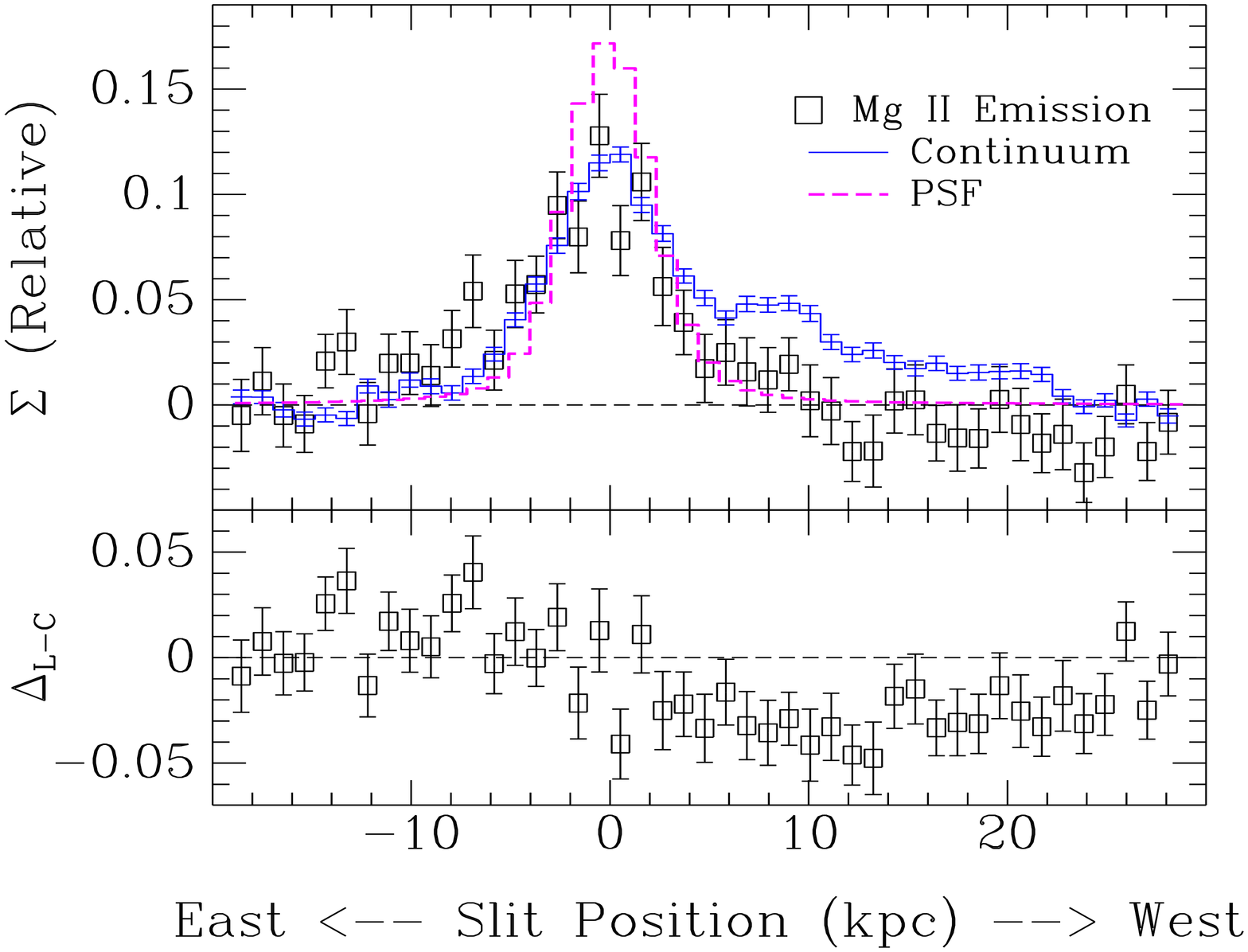}
                \hfill}
 \hbox{\hfill \includegraphics[height=8cm,angle=-90,trim=0 0 0 0]{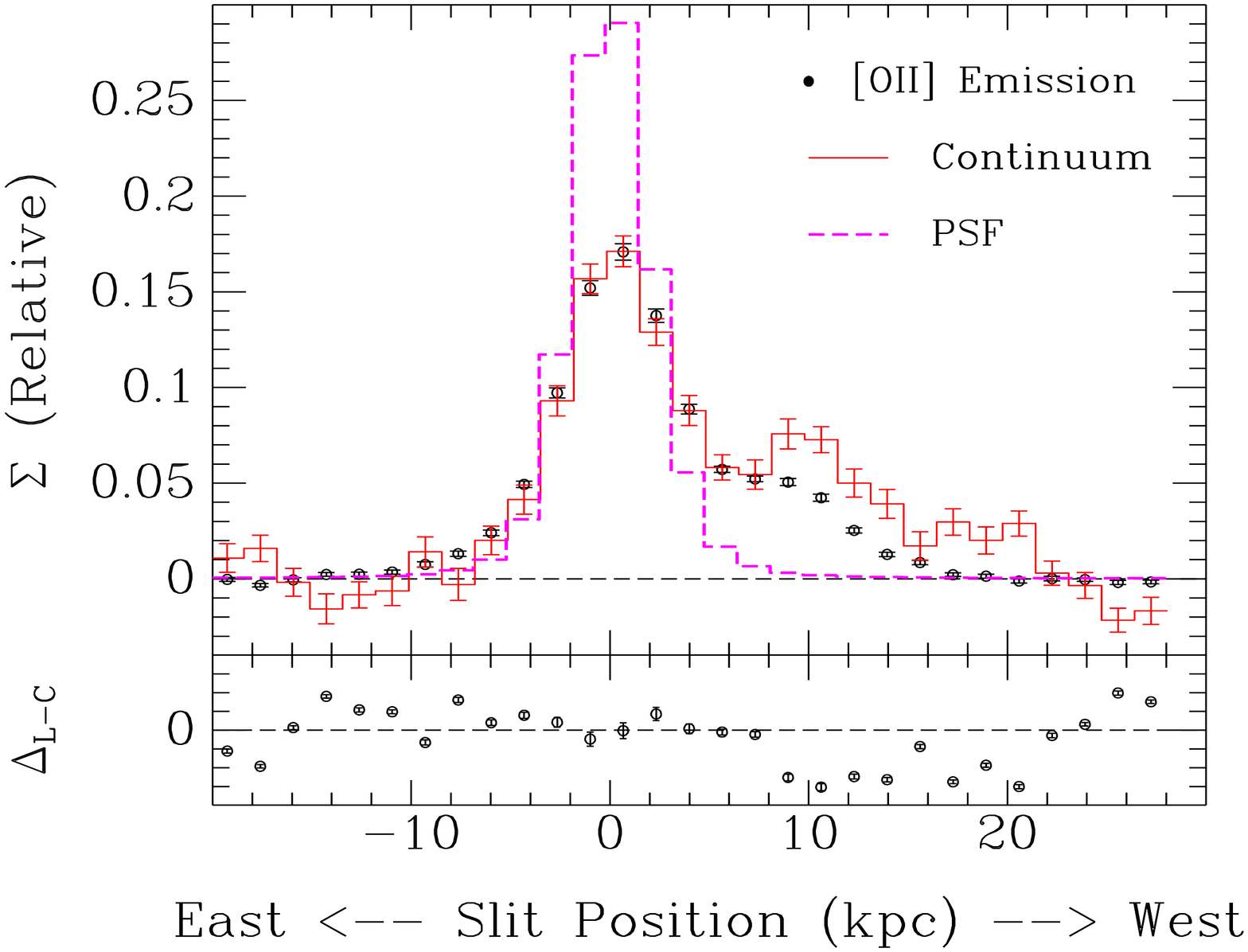}
                \hfill}
          \caption{\footnotesize
            Surface brightness profiles across 32016857. 
            The line emission is normalized to unity. To facilitate comparison to the
            emission line profile, the continuum has been scaled to match
            the maximum line emission. The residuals in the bottom panels
            show the difference of these
            scaled line and continuum profiles along the slit.
            The point spread function (psf) was measured from a stellar spectrum
            and has been normalized to unity area.
            {\it Top:} The \mgII\ doublet emission is spatially extended to the east, 
            as can be seen from the comparison to the stellar profile. At -10.6~kpc,
            the average \mgII\ surface brightness over one resolution element (6 pixels) 
            is significantly ($ 3.1\sigma $) above the background. The continuum emission 
            is also extended towards the east. In the lower panel, which shows the difference
            between the normalized line and continuum profiles, the line emission is 
            marginally ($2.1 \sigma $) more extended than the continuum emission.
            {\it Bottom:} In contrast, the [\ion{O}{2}] surface brightness profile
            is not spatially resolved to the east but is well resolved to the west. 
            Since the spatial distribution of the \mgII\ emission does follow
            that of the [\ion{O}{2}] emission, the former clearly does not come
            directly from \ion{H}{2} regions.
}
 \label{fig:emission_profile} \end{figure}

Figure~\ref{fig:emission_profile} compares the emission profiles along
the slit. The continuum emission, rather than showing the symmetric profile of
an inclined disk, shows an extended feature to the west with strong [\ion{O}{2}]
but no \mgII\ emission. The [\ion{O}{2}] emission is unresolved to 
the east. In contrast to this nebular line emission, the \mgII\ surface brightness
profile is spatially extended to the east but unresolved to the west. 
The opposite directions of the extended [\ion{O}{2}] and \mgII\ emission along the slit suggest
distinct physical origins.

The \mgII\ surface brightness averaged over a spatial resolution element (six pixels) 
10.6~kpc eastward of the galaxy is 3.1 standard deviations above the background. The
continuum emission is also spatially extended to the east.  Averaged over the same
spatial resolution element at 10.6~kpc, the line emission is marginally, $2.1 \sigma$,
more significant than the continuum emission indicating the line emission
is likely more extended than the continuum emission. The \mgII\ emission is not
spatially resolved to the west.

To estimate the radial extent of the \mgII\ emission, we introduce a simple model
for the surface brightness profile measured along the slit inspired by, but different
in detail from, the approach used by \cite{Rubin:2011p1660} to model the galaxy TKRS~4389. The
emission model includes an \ion{H}{2} region component and an extended component. 
The normalized [\ion{O}{2}] surface brightness profile determines the shape of
the \ion{H}{2} component. Since the observed \mgII\ profile is spatially resolved
to the east but not to the west, we require an asymmetric extended component. We
adopt a semi-circular emission region of constant surface brightness. The
radius, $R_C$, of the extended emission describes the projected distance from the center of 
32016857, and the straight-side of the semi-circle is perpendicular to the slit. This
2D model of the surface brightness profile is convolved with a Gaussian model
of the atmospheric seeing, $FWHM = 0\farcs8$ and then ``observed'' through a 1\farcs2
slit. In analogy with \lya\ radiative transfer, photons scattered off the far side of the 
outflow escape as redshifted emission.

\begin{figure}[t]
 \hbox{\hfill \includegraphics[height=8cm,angle=-90,trim=0 0 0 0]{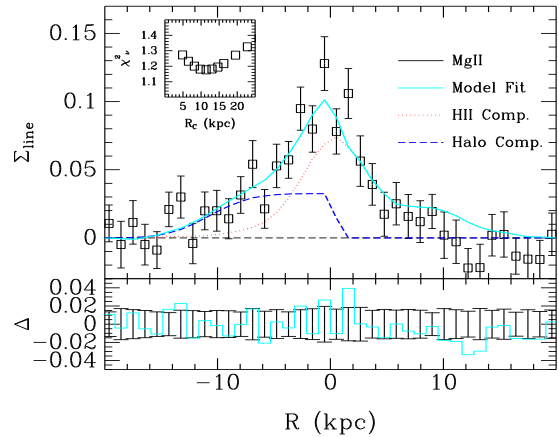}
                \hfill}
          \caption{\footnotesize
            The normalized \mgII\ surface brightness profile with fitted model (cyan line).
            This particular model, while not necessarily unique, provides an acceptable
            description of the \mgII\ surface brightness with fit statistic $\chi^2_{\nu} = 1.18$.
            The model is a linear combination of the emission from an HII region component (red,
            dotted line) and an extended halo (blue, dashed line). The normalized [\ion{O}{2}] 
            surface brightness profile
            defines the HII region component. The extended component is modeled by a semi-circular
            region of radius $R_C$ and constant surface brightness.  The best-fit radius for this 
            scattered component is $R_C = 11.4$~kpc, and the spacing between models is 0.3~kpc.
            The lower panel shows that the errors in the data values are comparable in magnitude
            to the the residuals from the fit (data minus model).
            The inset, however, shows the reduced chi-square values over a large range in $R_C$.
            Since the range in $R_C$ over which $\chi^2$ increases by 1 from its minimum
            value at 11.4~kpc is quite large, models with halo radii considerably smaller or larger 
            than 11.4~kpc are not ruled out.  
             }
 \label{fig:model_err} \end{figure}

We explore 14 values of $R_C$ from 4.9 to 35.9 kpc, corresponding to an angular extent
of 0\farcs621 to 4\farcs554 along the slit. For each model of the extended emission, we fit the \mgII\ surface
brightness profile with a linear combination of the \ion{H}{2} and extended components.
\fig~\ref{fig:model_err} shows the minimum residuals at $R = 11.4$~kpc, where the
separation between models is about 0.34~kpc. The extended component and the \ion{H}{2} 
component determine the net surface brightness profile to the east and west, respectively,
in \fig~\ref{fig:model_err}.  For the best-fit model, the eigenvalues of the two components 
are similar, and the extended component contributes 46\% of the total \mgII\ flux. While the
residuals show that this model provides a reasonable description of the data, it does not
tightly constrain the radius of the scattering halos. Smaller halos cannot be excluded with
high confidence due to the profile smearing caused by atmospheric turbulence, and this
blurring also means the reduced chi-square values increase slowly with increasing halo radius.

\section{Galaxies with Halos of Scattered Emission} \label{sec:halos_general}

Spectra of about 20\% of star-forming galaxies at $z \sim 1$ show resonance absorption
at large ($\sgreat 100$\kms) blueshifts relative to nebular and stellar lines 
\citep{Martin:2012}. Since the properties of star-forming galaxies with and without
blueshifted absorption do not differ very significantly, the absence of blueshifted
resonance absorption in some spectra has been attributed to the collimation of the outflows.
A bipolar flow, for example, produces a blueshift in the spectrum when the orientation of
the galactic disk is face-on and the starlight passes through the near side of the outflow cone. This
interpretation of the blueshifted fraction suggests outflows are a generic property of star-forming galaxies at
intermediate redshift and raises the question of why scattered emission is not detected more 
frequently.

Many properties of 32016857 and TKRS~4389 \citep{Rubin:2011p1660}, 
the only other distant galaxy detected in extended \mgII\ emission, are similar.
In particular, 
       \cite{Rubin:2010p808} suggest that TKRS~4389 is exceptionally 
       bright for its redshift of $z = 0.6943$.  It is slightly bluer and more luminous than 
       32016857 and has about three times as much stellar mass.
       A rough estimate for the SFR is $\approx 80$\msunyr. 
       The deepest parts of the \feII\ absorption 
       troughs are blueshifted roughly 200\kms.
       An HST image of 
       TKRS~4389 shows signs of recent merger activity. \cite{Rubin:2010p808} cite 
       the weak [\ion{Ne}{5}] emission, $W_r = -0.84 \pm 0.1$\AA, as evidence that 
       TKRS~4389, in contrast to 32016857, hosts a low-luminosity active galactic nucleus (AGN).

The very blue colors of 32016857 and TKRS 4389 are typical of \mgII\ emitters, which have
lower masses, bluer colors (i.e., less reddening), and higher sSFR than the average star-forming
galaxy at intermediate redshifts \citep[Rubin et al., in prep]{Erb:2012p26,Martin:2012}. 
However, these two galaxies are more luminous than the average \mgII\ emitter; their blue luminosities 
fall in the highest tertile of the \cite{Martin:2012} sample.  In spite of  the high luminosity of 
32016857, its stellar mass falls in the lowest tertile of our sample. The mass of TKRS 4389 is 
significantly higher, near the upper bound of our middle tertile. Galaxies that are both as luminous 
and as blue as 32016857 are rare. Only 9 of 208 galaxies in the parent sample have
$U - B < 0.459$ and $M_B -5 \log h_{70} < -21.3 $. Three of these 9 galaxies are \mgII\ 
emitters by the definition of \cite{Kornei:2013}, but only the 32016857 spectrum shows spatially 
extended \mgII\ emission.

Many factors, including the size of the spectroscopic aperture, the solid angle of the outflow and 
spectroscopic aperture, the amount of interstellar gas and dust, and the orientation of the galaxy and 
wind affect whether an outflow can be detected in resonance emission. Accordingly,
it is unclear whether the resonance photons only escape from these galaxies or whether
the emission is otherwise spread over too large a solid angle to detect it through a
narrow slit. Drawing from our observation that both 32016857 and TKRS 4389 have uncommonly
large luminosities, mapping \mgII\ emission in and around galaxies like these nine may be the most 
direct way to determine whether scattered halo emission is a generic property of intermediate
redshift galaxies. While this could be accomplished with LRIS using additional longslit 
position angles, the next generation of optical spectrographs with large integral field units,
particularly KCWI at Keck and MUSE at VLT, will be well positioned to characterize the halo
emission. 

With the aim of better informing these future investigations, we describe in Section~\ref{sec:mg2} 
additional highlights from our investigation of the spatial extent of \mgII\ emission in deep, LRIS spectra. From the union of a color selected sub-sample and the sample selected by  \mgII\ emission
in the integrated spectra, we identified two more (for a total of three) examples of extended 
\mgII\ emission which we discuss in this section. We also
describe a noteworthy example of extended \mgII\ absorption with unresolved emission in 22028686.
Finally, in Section~\ref{sec:fe2}, we remark on the absence of \feII\ $\lambda 2383$ emission.
We return to 32016857 in Section~\ref{sec:discussion} where we discuss the implications
of scattered \mgII\ emission for mass-loss rates. The reader primarily interested in
the physical properties of outflows rather than the challenge of detecting scattered
emission should proceed directly to Section~\ref{sec:discussion}.

\subsection{Additional Examples of Spatially Resolved \mgII\ Structure} \label{sec:mg2}

The bluest fifth of the \cite{Martin:2012} sample contains 42 galaxies with $U - B < 0.459$
and is of particular interest since the \mgII\ emitters tend to have blue colors \citep{Martin:2012,
Kornei:2013}. 
All but 4 of these spectra cover the \mgII\ doublet, yet only 10 of these 38 show prominent
\mgII\ emission in integrated spectra. Among this subsample, we found only two 
examples of spatially extended emission in the \mgII\ doublet. 
In addition to 32016857, the \mgII\ emission in our spectrum of 32010773 is spatially
extended.

\begin{figure*}[t]
 \hbox{\hfill \includegraphics[height=18cm,angle=-90,trim=0 0 30 0]{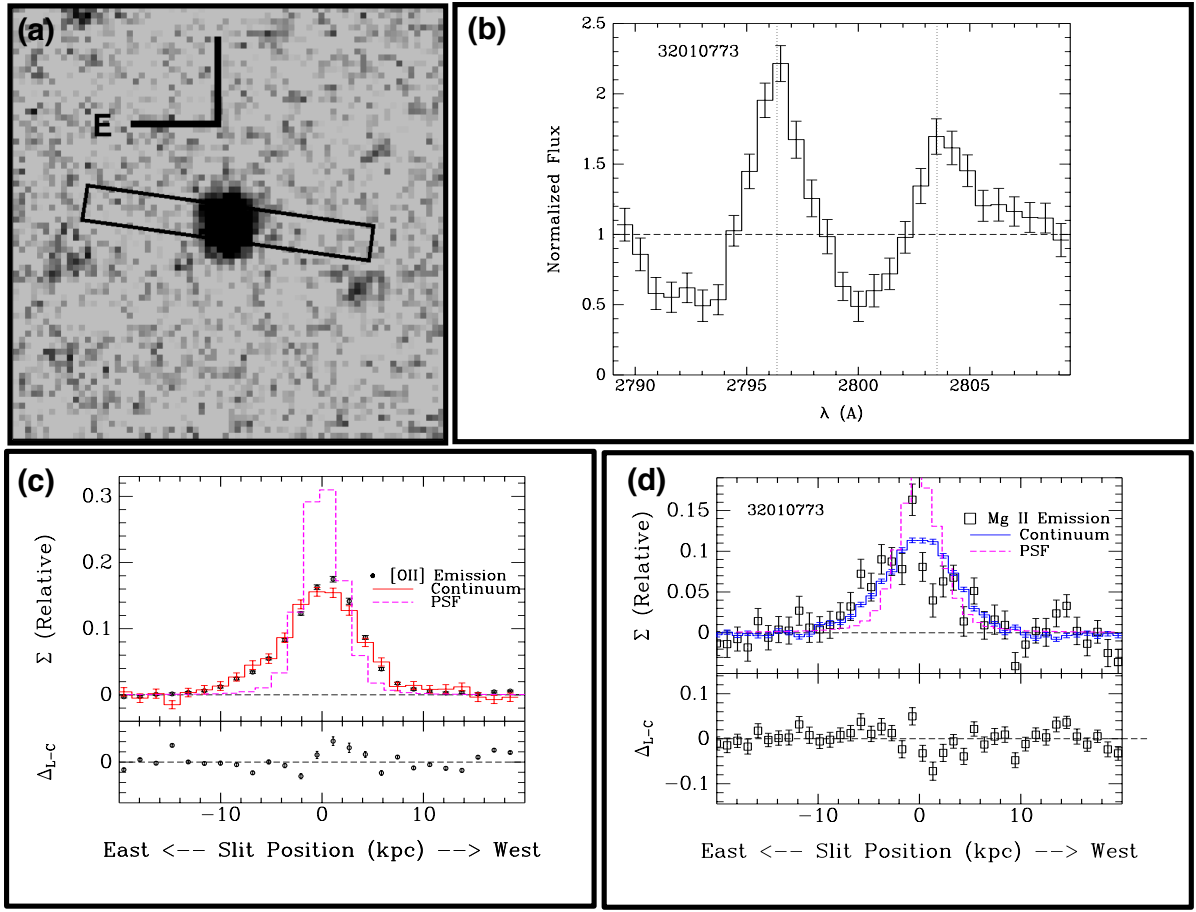}  \hfill}
          \caption{\footnotesize
Observations of 32010773.
(a) The galaxy, shown in a CFHT R-band image, was observed with LRIS
through a 1\farcs2 by 10\farcs0 slitlet at a position angle of 82.0\deg\ (on mask msc32aa as described
in Table 1 of \cite{Martin:2012}).
(b) The blue LRIS spectrum reveals a P-Cygni \mgII\ profile.
(c) Normalized surface brightness profiles of the [\ion{O}{2}] emission the nearby continuum 
are spatially extended relative to the point-spread function.
(d) Normalized surface brightness profiles of the \mgII\ emission is partially resolved but
no more spatially extended than the blue continuum emission.
The Gaussian FWHM fitted to the \mgII\ surface brightness profile, $1\farcs25 \pm 0\farcs11$,
is not significantly larger than the FWHM of the blue continuum profile, $1\farcs14 \pm 0\farcs02$.
The fitted centroid of the \mgII\ emission profile is 1.2 pixels (0\farcs16) east of 
the continuum centroid, a $2.9 \sigma$ discrepancy in position.
}
 \label{fig:32010773} \end{figure*}

Figure~\ref{fig:32010773} shows the surface brightness profiles of the \mgII\ emission (panel~d), blue continuum,
[\ion{O}{2}] emission (panel c), and red continuum along the slit crossing 32010773. In contrast to
the surface brightness profiles measured across 32016857, 
the \mgII\ emission across 32010773 
is no more extended than the blue continuum emission. In the continuum-subtracted,
2D spectra, the velocity gradients of the \mgII\ and [\ion{O}{2}] emission along the slit
do not obviously differ, so the \mgII\ emission may come directly from \ion{H}{2} regions. Our data
do not require (nor do they rule out) scattered \mgII\ emission. This galaxy does clearly
have an outflow; the large blueshift of the \feII\ resonance lines, $ V_1 = -198 \pm 46$\kms,
is quite significant. The color of 32010773, $U - B = 0.23$ is among the bluest in the sample;
but the luminosity of 32010773, $M_B - 5 \log h_{70} = -20.68$, is much lower than the luminosity
of either 32016857 or TKRS 4389. The absence of resonance emission that is clearly extended relative to the continuum
around 32010773 is consistent with the conjecture that detection of a scattering halo requires 
especially luminous galaxies.  

\begin{figure*}[t]
 \hbox{\hfill \includegraphics[height=18cm,angle=-90,trim=0 0 30 0]{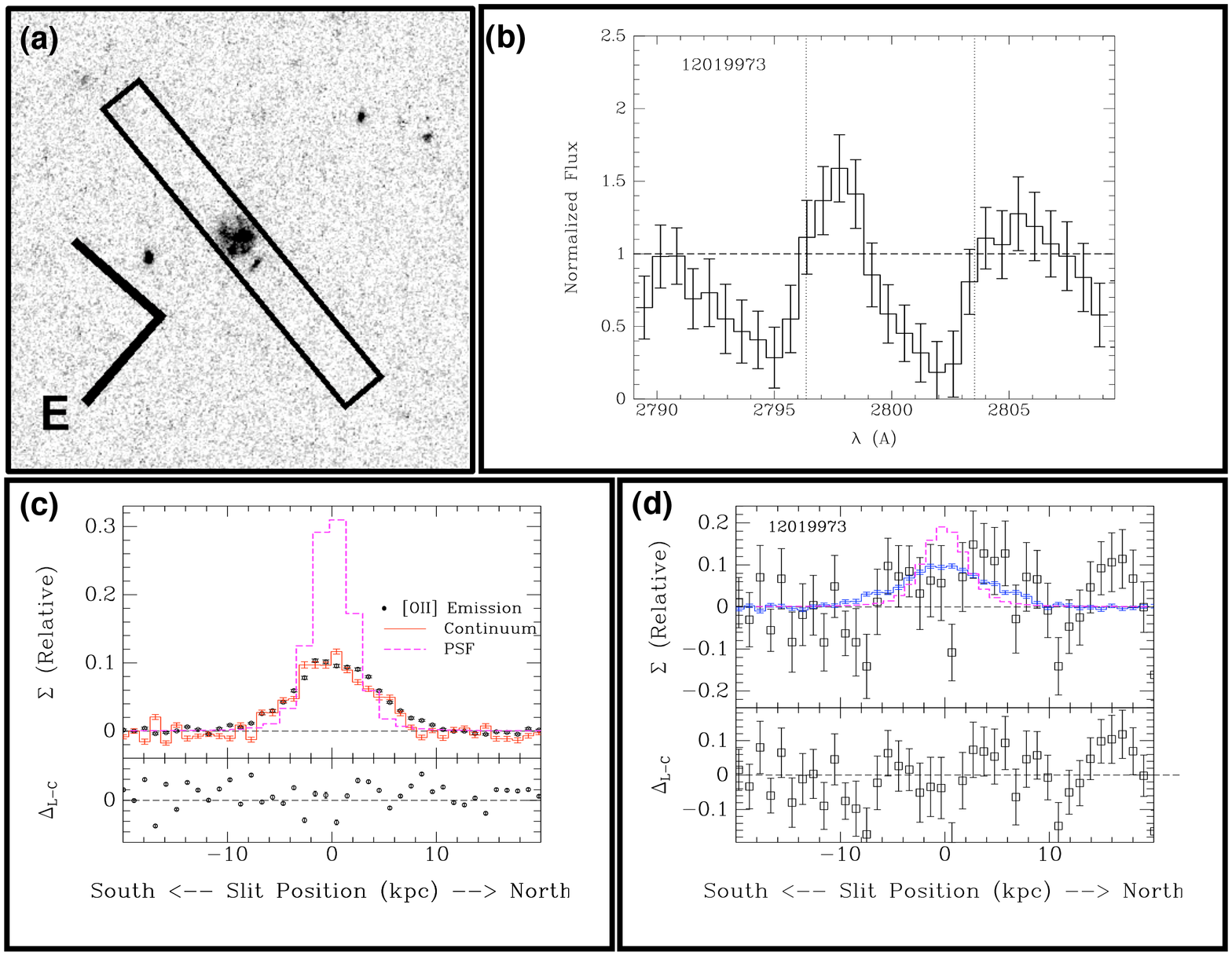}    \hfill}
          \caption{\footnotesize
Observations of 12019973.
(a) The galaxy is shown on the V band HST image described in \cite{Kornei:2012p135}. We obtained
the LRIS spectrum through a 1\farcs2 by 10\farcs0 slitlet at a position angle of 170.0\deg\ (on mask msc12ee as 
described in Table 1 of \cite{Martin:2012}.)
(b) The blue LRIS spectrum reveals a P-Cygni \mgII\ profile.
(c) Along the slit, the strong [\ion{O}{2}] emission is slightly more extended than our model of
the psf but not significantly more extended than the red continuum emission.
(d) The \mgII\ surface brightness profile exemplifies the low S/N ratio typical of \mgII\ emission.
}
 \label{fig:12019973} \end{figure*}

In addition to these 10 \mgII\ emitters with very blue color, \cite{Kornei:2013} identified 
12 more \mgII\ emitters  in the \cite{Martin:2012} sample with $U-B$ redder than 0.459. 
Among these redder \mgII\ emitters,
we found spatially extended \mgII\ and continuum emission across 12019973.
Figure~\ref{fig:12019973} shows the variation in the \mgII\ and [\ion{O}{2}] emission along the slit. 
Across 12019973, much like 32010773, the  surface brightness profile of 
the line emission is no more extended than the continuum emission, neither proving nor ruling out
scattering by an outflow. We note that the Doppler shift of the \feII\ resonance lines,
$V_1 = -59 \pm 28$\kms, provides marginal evidence for an outflow from this galaxy. The color, $U - B = 0.539$, 
and stellar mass, $\log (M/ \msun) = 10.15$, of 12019973 are typical of the sample. This galaxy has a relatively 
low luminosity, $M_B - 5 \log h_{70} = -20.57$, which may contribute to the faintness of any extended halo.

\begin{figure*}[t]
 \hbox{\hfill \includegraphics[height=18cm,angle=-90,trim=0 0 100 0]{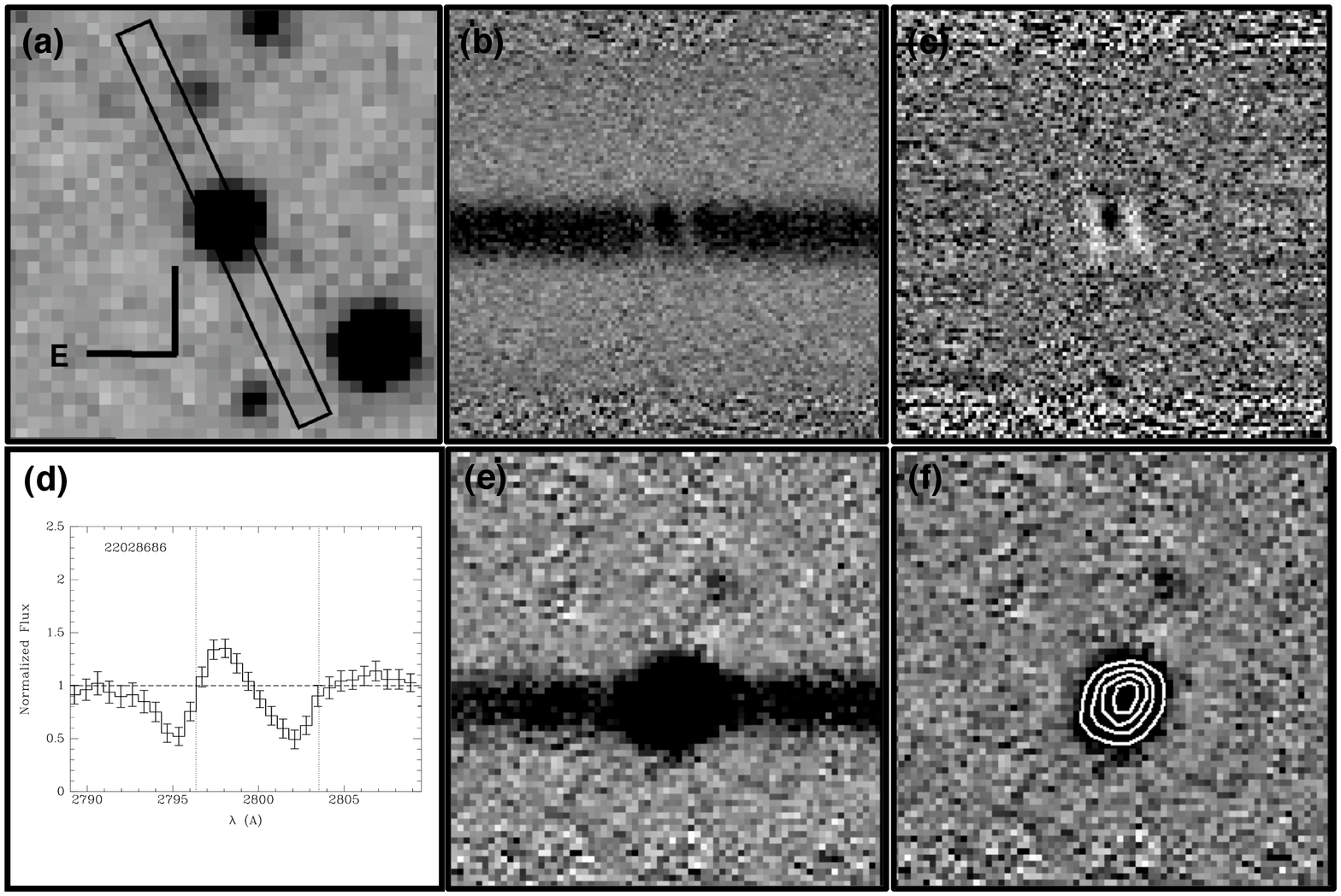}    \hfill}
          \caption{\footnotesize
Observations of 22028686.
(a) CFHT R-band image. The galaxy was observed with LRIS through a 1\farcs2 by 15\farcs0 slitlet at 
a position angle of 25.0\deg\ on mask msc22bb as described in Table 1 of \cite{Martin:2012}. 
(b) A 140 \AA\ wide segment of the blue LRIS spectrum centered on the \mgII\ doublet. 
Wavelength increases to the right, and northeast is up. The prominent
absorption lines exhibit a velocity gradient across the galaxy; the centroid shifts blueward from the
systemic velocity on the southwest side to  approximately $-120 \pm 30$\kms\ 1\farcs08 (8.31~kpc)
to the northeast.
(c) Continuum subtracted blue spectrum. The \mgII\ emission component is spatially unresolved.
The continuum emission subtends a larger angle than the resonance emission.
(d) Integrated \mgII\ profile. The P-Cygni shape of the line profile, blueshifted absorption and 
redshifted emission, is typical of our \mgII\ emitter sample.
(e) A 90 \AA\ wide segment of the red LRIS spectrum. The blended [\ion{O}{2}] doublet is spatially extended. 
Wavelength increases to the right and northeast is up.
Linearly spaced contours show a slight tilt of the nebular emission in the opposite sense of the velocity 
gradient apparent in panel (b). 
(f) Continuum subtracted red spectrum. The faint, serendipitous emission line near the northeastern 
end of the slit is 4\farcs44 northeast of the peak continuum emission along the slit, and we
identify it as [\ion{O}{2}] emission from the $R = 24.8$ DEEP2 galaxy 22028464, which is visible 4\farcs66
northeast of 22028686 in panel (a).
}
 \label{fig:22028686} \end{figure*}

Figure~\ref{fig:22028686} shows the 2D spectra of another \mgII\ emitter, 22028686, a star-forming
galaxy with $U-B > 0.459$. Although the \mgII\ emission is not spatially resolved, the resonance
absorption troughs extend well beyond the emission region.
The \mgII\ absorption lines are tilted in the opposite sense of the velocity gradient indicated
by the [\ion{O}{2}] Doppler shifts along the slit. Since the [\ion{O}{2}] emission 
defines the systemic redshift, the redshifted and blueshifted [\ion{O}{2}] emission to
the northeast and southwest of the galactic center reveal either the projected rotation of the galaxy
or the orbital motion of two merging galaxies.
The centroid of the \mgII\ absorption to the southwest of the galaxy is consistent with
the velocity of the blueshifted side of the galaxy. It is the even larger blueshift
of the \mgII\ absorption centroid to the northeast that makes the absorption lines tilt
in the opposite direction of the [\ion{O}{2}] emission. Where the \mgII\ absorption is blueshifted
on the northeastern side of the galaxy, we detect strong resonance emission at redshifted
velocities. No resonance emission is detected along the southwestern half of the slit. 
We note that the \feII\ absorption in the integrated spectrum does not show a net Doppler shift, 
$V_1 = -1.9 \pm 9.2$~km~s$^{-1}$. Much of the shift in the centroid of the \mgII\ 
absorption may therefore be caused by variations in emission filling along the slit, and
the spatial gradient in the absorption velocity likely does not reflect galactic rotation.

The detection of \mgII\ emission from just a portion of the absorbing region in 22028686
may help illuminate the physical conditions required for the resonance emission to escape the galaxy.
To gain this insight, we review studies which have mapped \ion{Na}{1} resonance absorption across galaxies.
As illustrated for example by spectral mapping of  NGC 1808 \citep{Phillips:1993} 
local ULIRGs \citep{Martin:2006}, and Mrk~231 \citep{Rupke:2005c,Rupke:2011p27}, 
the absorption from outflows is not redshifted
anywhere across the galaxy because the absorbing gas must lie in front of the continuum source.
Across local ULIRGs,  the \ion{Na}{1} velocity gradient is very similar to the slope of the rotation curve, but
the absorption trough is highly blueshifted everywhere along the slit. This situation would be expected if 
the outflow was launched from a rotating disk, and we might expect a small velocity gradient in
the absorption Doppler shift (not seen) comparable to the tilt of the nebular emission lines. In spite
of the very limited spatial resolution across 22028686, we speculate that the geometry might instead
mimic NGC~1808 where differences in extinction shape the resonance line profile of the outflow.
As illustrated in Figures~1 and 4 of \cite{Phillips:1993}, the blueshifted lobe of the outflow is 
detected in absorption across the inclined disk. Resonance emission is detected from the
far side of the outflow where it pokes out from behind the disk but plumes of dust visible 
on the near side of the outflow at the location of the slit presumably destroy the resonance 
photons. Hence the implication from the 22028686 spectrum seems to be that the 
sightline towards the absorbing gas on the northeast side encounters less dust than that 
towards the southwest side of the slit. Higher resolution imaging could confirm the presence
of a disk, determine the disk orientation, and perhaps reveal the locations of dust filaments
in this galaxy.

In summary, among 145 spectra with coverage of the \mgII\ doublet, including
50 galaxies with either prominent \mgII\ emission or very blue $U-B$ color, we
find only three examples of spatially resolved \mgII\ emission:  32016857, 32010773,
and 12019973.  Our analysis of 32016857 doubles the number of known galaxies with \mgII\ 
emission extending beyond the stellar continuum. With only two examples where the
extended emission must be scattered, we can only begin to speculate about the conditions 
producing scattered  \mgII\ halo emission. However, since the unique properties of 32016857 and 
TKRS 4389 are their high luminosities (for their blue color), we suggest that large luminosities 
are essential to raise the surface brightness of the scattered emission to detectable levels. 
Further support for the idea that most scattering halos are below current detection limits
comes from stacking the surface brightness profiles of $1 < z < 2$ \mgII - emitting galaxies;  
their \mgII\ emission is marginally more extended than the near-UV continuum
at radii of 0\farcs8 \citep{Erb:2012p26}.

\subsection{Absence of Extended Emission in \feII\ Lines} \label{sec:fe2}

Our spectrum of 32016857 shows prominent fluorescent emission in Figures
\ref{fig:intro} and \ref{fig:vc_fit}.
We did not find extended emission in resonance  lines of \feII\
or fluorescent \feII$^*$ lines nor did \cite{Rubin:2011p1660} in
their TKRS~4389 spectrum. The \feII\ $\lambda 2383$  absorption is 
optically thick in most galaxy spectra; under LS coupling, the
only permitted decay from the upper energy level, here the excited $z^6F^0_{11/2}$ 
state of $Fe^+$, is to the ground $a^6D_{9/2}$ state.  The absence of \feII\ $\lambda 2383$ 
emission in spectra with prominent \mgII\ $\lambda 2803$ emission is at first
surprising considering the similar oscillator strengths of the $\lambda 2383$
and $\lambda 2796$ transitions and the (only) 20\% larger cosmic abundance of 
magnesium (relative to iron). 

Several factors may contribute to the paucity of \feII\ $\lambda 2383$ emission.
A greater depletion of iron (relative to magnesium) onto 
grains, for example, may reduce the \feII\ $\lambda 2383$ optical depth relative
\mgII\ $\lambda 2803$. In the Milky Way, however,
the relative depletion of Fe and Mg in the halo clouds is much lower than the ratio
measured for disk clouds \citep{Savage:1996p1461}. Hence, we might 
anticipate a similar depletion of Fe and Mg onto grains in outflows.  
Alternatively, since iron is not a light element,
deviations from pure LS coupling might in principle allow decay to another 
term thereby reducing the resonance emission (Shull, pvt. comm.), 
but we find no observational evidence
for unidentified emission lines in support of this conjecture in the near-UV spectra. 
Third,as suggested recently by \cite{Erb:2012p26}, boosting the \mgII\ emission, rather 
than suppressing the \feII\ emission, offers another explanation.  Those authors show that 
photoionized gas produces more emission in \mgII\ than \feII, so emission from \ion{H}{2} 
regions may boost the intensity of the scattered \mgII\ radiation relative to the scattered 
\feII\  photons.  

The most important factor driving this discrepancy, however, appears
to be the ionization corrections.   \cite{Giavalisco:2011} show (see their Figure 14) that 
photoionization models of very low column density ($N_H \sim 10^{18}$\col) gas indicate
a large plausible range in the relative column density; $N_{MgII} / N_{FeII}$ increases 
from approximately unity ($\log U = -4$) to just over 100 at $\log U = -2$.  Even though the
first (7.646 eV, 7.870 eV) and second (15.035 eV, 16.18 eV) ionization potentials of 
Mg and Fe, respectively, are quite similar, the Fe is mainly in ${\rm Fe}^{+3}$ while the Mg
is in ${\rm Mg}^{++}$ in our photoionization models, which we describe in the next section.
This happens because it takes only 30.6~eV to ionize ${\rm Fe}^{++}$ to 
${\rm Fe}^{+3}$ (and 54.8~eV to make ${\rm Fe}^{+4}$) but 80~eV to remove another electron 
from ${\rm Mg}^{++}$.

\section{Discussion} \label{sec:discussion}

\begin{figure*}[t]
 \hbox{\hfill \includegraphics[height=18cm,angle=-90,trim=0 0 0 0]{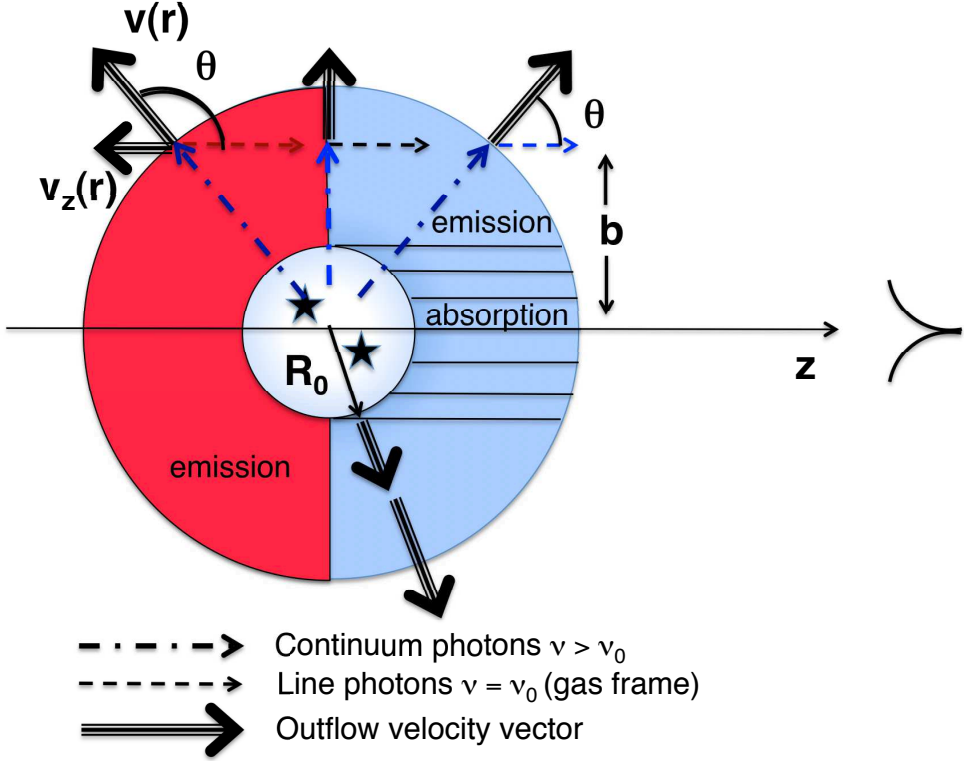}      \hfill}
          \caption{\footnotesize
Schematic diagram of a spherical outflow around a galaxy illustrates an
opening angle of $\Omega = 4\pi$ steradians, equivalent to a conical outflow
with $\theta_{cone} = \pi$. The regions colored red and blue denote redshifted
and blueshifted emission, respectively; and the hashed region denotes the location
of atoms producing blueshifted absorption. For a delta-function line profile at the resonance
$\nu_0$, atoms at each point along a sightline through the outflow absorb continuum 
photons at a unique frequency $\nu = \nu_0 (1 + v(r) / c)^{-1}$, where $r = \sqrt{b^2 + z^2}$
along a sightline at impact parameter $b$. The emitted line photons have Doppler shifts
corresponding to the range of projected outflow velocities, $v(r) \cos \theta$.  For a
constant velocity gradient, the density falls as $r^{-3}$, and we find the maximum Sobolev
optical depth along the sightline at $z = 0$.
}
 \label{fig:geometry} \end{figure*}

In this section we apply the idea of resonance scattering to the spatial extent of \mgII\
emission. We take as a model for the outflow a purely radial flow as illustrated in 
Figure~\ref{fig:geometry}. 
The radial component of the velocity is critical since the photons 
emitted by stars and \ion{H}{2} regions emanate from smaller radii. 
Atoms in the circumgalactic medium (CGM) scatter photons 
at the resonance frequency in their rest frame. 
Outflowing gas absorbs photons blueward of resonance. 
The Doppler shift of the absorption trough directly reflects the velocity of the
gas located between the galaxy and the observer.
The redshifted emission from the far side
of the outflow may escape and produce a redshifted emission line. 
The size of the scattering halo indicates the radius where the optical depth
along a radial ray is approximately unity.

The clouds accelerate near the galaxy due to the radial decline
in both thermal gas pressure and radiation pressure. The continuity
equation gives the mass flux,
\begin{eqnarray}
\dot{M}(r) = \Omega f_c \rho(r) v(r) r^2,
\label{eqn:conserve} \end{eqnarray}
at radius $r$ through the solid angle $\Omega$ subtended by the wind. 
The cloud covering fraction, $f_c$, describes the fraction of the outflow
solid angle covered by low-ionization clouds (as seen from
the galaxy).

Defining $\bar{m}_{ion}$ as the mass in the warm outflow per ion,
the column density between the observer and the launch radius, $R_0$, is 
\begin{eqnarray}
N_{ion} = \int_{R_0}^{\infty} \rho(r) / \bar{m}_{ion}  dr.
\label{eqn:N} \end{eqnarray}
Solving Eqn.~\ref{eqn:conserve} for $\rho(r)$, we see that the
outflow density decreases with the inverse-square of the radius
for the simple case of a constant velocity outflow with no
mass-loading beyond the launch radius. For this example,
the integral in Eqn.~\ref{eqn:N} yields a column density that
declines inversely with the launch radius $R_0$. Accurately
measuring the ionic column density will still leave the 
mass-loss rate uncertain at the level of the plausible
range of launch radii as well as the total mass per ion.
Increases in column density strengthen the absorption trough 
significantly in the optically thin limit; but for saturated 
lines like the \mgII\ doublet, the range in gas velocity 
along the sightline largely determines the equivalent width 
of the intrinsic absorption troughs. The measured absorption
equivalent width therefore provides only a lower limit on
the ionic column density. The absorption troughs imprinted
by outflows in galaxy spectra provide no constraint on the 
location of the gas along the sightline.

Resolving scattered line emission constrains the location and density of
the gas flow. The scattering atoms must lie at radii at least as large as the 
projected impact parameter. The scattering optical depth scales with the
gas density for a given model (or measurement) of the velocity gradient in
the outflow. 

In this section, we first we estimate the ionic density and 
radius at the scattering surface in 32016857 and TKRS~4389.  Then we discuss the limits
on the relationship between the total gas density and the ionic density.
Finally, we illustrate the implications for the mass flux in low-ionization gas.

\subsection{Density of $Mg^+$ in Outflowing Gas} \label{sec:density}

We model the outflows with a radial velocity field. For 32016857, we picture 
a scenario where our slit intersects the outflow cone on only the west side of 
galaxy. In  TKRS~4389, in contrast, the longslit observation resolves \mgII\ emission 
on both sides of the slit \citep{Rubin:2011p1660}.
A simple model with a constant surface brightness halo suggests the impact
parameter of the scattering surface is $8.25 < b ({\rm kpc}) < 13$ for
TKRS~4389 \citep{Rubin:2011p1660} and roughly 11.4~kpc from 32016857.

The interaction region where a resonance transition such as \mgII\ $\lambda 2796$ 
can be scattered depends on the line profile. To simplify the calculation,
we consider a delta function line profile centered on the resonance frequency $\nu_0$ 
(in the reference frame of the bulk flow). Photons emitted by the starburst at frequency $\nu$
propagate radially until they encounter a gas parcel with radial velocity $v(r_S) = 
c (\nu - \nu_0) / \nu_0$.  We call the interaction spot the Sobolev point, $r_S$. 
Provided velocity $v$, $\mu$, and the radial velocity gradient change slowly around 
$r_S$, the Sobolev expression \citep{Sobolev:1960,Lamers:1999}
describes the scattering optical depth, 
\begin{equation}
\tau^S_{\nu_0} = \frac{\pi e^2}{m_e c} f \lambda_0 n_i \nonumber \\
\left ( (1 - \mu^2) \frac{v(r)}{r} +  \mu^2 \frac{d v}{d r} \right)^{-1},
\label{eqn:sobolev} \end{equation}
where $f$ represents the transition oscillator 
strength.\footnote{Note that the stimulated emission term has been 
     dropped from Eqn.~\ref{eqn:sobolev} due to the low gas 
     densities in galactic outflows.}
The incident photons and the velocity of the bulk flow are both radial, so the cosine 
of the angle $\theta$ between them takes the value $\mu = \cos \theta = 1$. 
The frequency of the photons absorbed from the starburst varies only with $v(r)$ for a 
spherical outflow.

With $\mu = 1$ in Eqn.~\ref{eqn:sobolev}, the Sobolev optical depth for the stronger \mgII\ 
transition is given by
\begin{eqnarray}
  \tau_{2796}^S (r) = 4.57 \times 10^{-7} {\rm ~cm}^{3} {\rm ~s}^{-1} n_{{\rm Mg}^+}(r) \left 
   \vert \frac{dv }{dr} \right \vert^{-1} .
   \label{eqn:tau} \end{eqnarray}
At some radius along the sightline at impact parameter $b \approx 11.4$~kpc, 
we expect the scattering optical depth for the starburst photons to be approximately unity. 
We estimate the gas density
by setting $\tau_{2796}^S(r) \approx 1$ and obtain 
\begin{eqnarray}
n_{{\rm Mg}^+} \approx 1.4 \times 10^{-9} {\rm ~cm}^{-3} \left \vert \frac{dv}{230 {\rm ~km~s}^{-1}} 
\frac{11.4 {\rm ~kpc}}{dr} \right \vert_{r_S}.
\label{eqn:nMg} \end{eqnarray}
If the sensitivity of our spectrum, rather than the radius of the last scattering surface,
determines the observed extent of the scattered emission along the slit, then the optical depth 
(and the inferred density) would be higher than this estimate. 
However, we think modeling the spatial
extent with the last scattering surface is roughly correct
because the emission is not optically thick at $b \approx 11.4$~kpc; 
at the largest impact parameters where \mgII\ emission is detected, only $\lambda 2796$ is 
                detected, consistent with a two-to-one flux ratio and an
                optical depth in $\lambda 2803$ less than unity.
The gas density estimated from Eqn.~\ref{eqn:nMg} depends on the velocity gradient
at the Sobolev point, and this radius is not uniquely determined from the impact 
parameter $b$ alone.

Fig.~\ref{fig:geometry} illustrates the geometry for the outgoing, scattered
photons at impact parameter $b$. Along a sightline, the angle $\theta$ ranges from 
90\deg\ where the flow is perpendicular to the sightline to a minimum, $\theta_{min}$, 
and maximum $\theta_{max} \equiv 180\deg - \theta_{min}$, constrained by the Doppler shifts of the
absorption and emission lines. The Doppler shift of a scattered photon
reflects the projected velocity component, $v_z$, and will be most blueshifted
(redshifted) as $\theta_{min} \rightarrow 0\deg$ ($\theta_{max} \rightarrow 180\deg$). 
These extrema mark the scattering that occurs at the largest radius along a sightline;
and we infer the largest (smallest) radius from the lower (upper) limits on $\theta_{min}$.
The error bars on the angle come from considering the projection of the 
radial velocity (measured in absorption) onto to the line-of-sight component 
measured at impact parameter $b$ in emission. For 32016857, we estimate $38\deg \le
\theta_{min} \le 75\deg$, where the spectral fitting constrains the most
blueshifted absorption ($(|V_{Dop}| + b)$) to  the range $185 - 383\kms$; and the 
the redshifted emission  ($V_z + \sigma$) extends to at least $ 100 - 145 \kms$.
The maximum radius from which we detect scattered emission is therefore between
$1.04 b \approx 12$~kpc and $1.61 b \approx 18$~kpc. 

We estimate an average velocity gradient of 230\kms\ over 12 to 18 kpc. Substituting
these values into Eqn.~\ref{eqn:nMg}, we infer an ionic density between
$8.9 \times 10^{-10} \rm {~cm}^{-3}$ for $r_S = 18$~kpc ($\theta_{min} = 38\deg$)
up to $1.3 \times 10^{-9} \rm {~cm}^{-3}$ for $r_S = 12$~kpc ($\theta_{min} = 75\deg$).
A specific form for the acceleration of the outflow, $v(r) = v_0 \sqrt{\ln r/R_0}$,
demonstrates how a more physical model might change our density estimate. The
acceleration of a momentum driven outflow begins rapidly and then slows down
with increasing distance. This model reaches a characteristic velocity $v_0$
at a radius a few times the launch radius, $ r = 2.72 R_0$.  Choosing $R_0$
such that $v(r_S) = v_0$ and $r_S = 2.72 R_0$, we obtain $dv/dr = 0.5  v_0 / r_S$. 
Placing the Sobolev point at $2.72 R_0$ for comparative purposes,
the estimated velocity gradient at $r_S$ takes a value half as large as the constant
acceleration limit; and our density estimate would decrease by a factor of two.

For TKRS~4389, \cite{Rubin:2011p1660} measured the Doppler shift of the low-ionization absorption 
lines and fit a radius for the scattering halo to the surface brightness profile of the \mgII\
emission. Substituting their values of $V_{Dop} \approx 200 - 300$\kms\ and 
$b \approx 8.25 - 12.4$~kpc into Eqn.~\ref{eqn:nMg}, we estimate $n_{Mg+} \approx 1.1 - 2.6 \times 
10^{-9}$\cm3.  Since the Sobolev radius could lie at radii as large as $r_S \approx 1.5b$, where
we have used the Doppler shift of the $\lambda 2803$ emission,  $v_z \approx 226$\kms\ in 
Table 1 of \cite{Rubin:2011p1660}, and $v = 300$\kms\ to estimate
$\theta_{min} = \cos^{-1}(v_z(r_S) / v(r_S)) \approx 41\deg$, we estimate a  lower limit on the 
ionic density of roughly $0.73 \times 10^{-9}$\cm3.
The slightly larger density, $n_{Mg+} \approx 7 \times 10^{-9}$\cm3, found
by \cite{Rubin:2011p1660} is consistent with substituting the -800\kms\ Doppler shift of the blue
wing on the \mgII\ profile into Eqn.~\ref{eqn:nMg} along with the minimum Sobolev radius.
These arguments constrain the ionic density at the Sobolev radius in the TKRS~4389 outflow
to the range $ 0.7 - 7 \times 10^{-9}$\cm3.

\subsection{Gas Density}

Given the ionic density $n_{{\rm Mg}^+}$, the total gas density scales inversely with the 
ionization fraction, metallicity, and dust depletion. Defining the ionization fraction
\begin{eqnarray}
\chi({\rm Mg^+}) \equiv n_{{\rm Mg^+}} / n_{{\rm Mg}}, 
\end{eqnarray}
the hydrogen number density is 
\begin{eqnarray}
n_{H} = \frac{n_{{\rm Mg^+}} }{\eta({\rm Mg}) d({\rm Mg}) \chi({\rm Mg^+}) }.
\end{eqnarray}
We scale to solar metallicity, $\eta({\rm Mg}) = 3.8 \times 10^{-5}$, which
is appropriate (to a factor of two or better) for a wind loaded with interstellar
gas from a galaxy with the stellar mass of the 32016857 \citep{Zahid:2011p137}. 
We chose a fiducial value for the depletion of {\rm Mg} onto grains
typical of clouds in the Milky Way disk, $d({\rm Mg}) \approx 6.3 \times 10^{-2}$
\citep{Savage:1996p1461},
and note that a lower depletion similar to clouds in the Milky Way halo 
($d \approx 0.28$) would lower our density estimate by a factor of four.

To estimate $\chi({\rm Mg^+})$, we start with a rough assumption, which
we relax later, namely that the bulk of the Mg ions are either ${\rm Mg}^+$ 
or ${\rm Mg}^{++}$. Then 
\begin{eqnarray}
\chi({\rm Mg}^+) \approx n_{{\rm Mg^+}} / n_{{\rm Mg^{++}}} 
= \frac{\alpha({\rm Mg^+})}{c\sigma({\rm Mg^+})} U^{-1},
\label{eqn:trick} \end{eqnarray}
where the ionization parameter $U$ is defined as
\begin{eqnarray}
U \equiv \frac{Q}{4 \pi r^2 c n_e},
\end{eqnarray}
and
\begin{eqnarray}
\epsilon_{Mg+} \approx \frac{\alpha_{Mg+}}{c \sigma_{Mg+}} \approx 4.7 \times 10^{-4}.
\end{eqnarray}
In the last expression, the recombination rate $\alpha_{Mg+}(T) \approx
3.5 \times 10^{-12} {\rm ~cm}^3 {\rm ~s}^{-1}$ at $T = 10^4$~K \citep{Shull:1982p95},
and the photoionization cross section $\sigma_{Mg+} \approx 2.5 \times 10^{-19} {\rm ~cm}^2$
\citep{Verner:1996p487}. For 32016857, we estimate the ionizing photon luminosity $Q \approx 1.3 
\times 10^{55} {\rm ~s}^{-1}$ from  the  \Ha\ luminosity
(see description in Table 1) and the relationship in 
\cite{Kennicutt:1998araa}.

Using Eqn.~\ref{eqn:trick}, we can solve for $n_{H} \approx n_e$
\begin{eqnarray}
n_{H} \approx \left [ 
             \frac {n_{{\rm Mg^+}} Q}{4 \pi c \epsilon_{\rm Mg^+} \eta({\rm Mg^+}), d({\rm Mg}) r^2}
               \right ]^{1/2}.
\label{eqn:density}  \end{eqnarray}
At the location where $\tau_{2796} \approx 1$, 
the fiducial values give a {\rm H} density 
\begin{eqnarray}
n_{H} = 0.30 {\rm ~cm}^{-3} 
\left ( \frac{n_{{\rm Mg}^+}}{10^{-9} {\rm ~cm}^{-3} }  \right )^{1/2}
\left ( \frac{11.4 {\rm ~kpc}}{r_S} \right ) \times \nonumber \\
\left ( \frac{6.3 \times 10^{-2}}{d({\rm Mg})} \right )^{1/2}
\left (\frac{3.8 \times 10^{-5}}{\eta({\rm Mg})} \right )^{1/2}
\left (\frac{Q}{10^{55} {\rm ~s}^{-1}} \right )^{1/2}.~~
\label{eqn:nH} \end{eqnarray}
The implied ionization parameter at the Sobolev point is
\begin{eqnarray}
U(r_S) \approx 7.6 \times 10^{-2}
\left ( \frac{Q}{10^{55} {\rm ~s}^{-1}} \right )
\left ( \frac{0.30 {\rm ~cm}^{-3}}{ n_e} \right ) \times \nonumber  \\
\left ( \frac{11.4 {\rm ~kpc}}{ r_S} \right )^2.
\end{eqnarray}
The resulting ionization fraction of ${\rm Mg}^+$,
\begin{eqnarray}
 \chi({\rm Mg}^+)  \approx 1.4 \times 10^{-3}
\left ( \frac{10^{55} {\rm ~s}^{-1}}{Q} \right )
\left ( \frac{ n_e}{0.3 {\rm ~cm}^{-3}} \right ) \times \nonumber  \\
\left ( \frac{ r_S}{11.4 {\rm ~kpc}} \right )^2,
\end{eqnarray}
indicates most of the Mg in the warm phase of this outflow
has been ionized to ${\rm Mg}^{++}$. For comparison, inspection of
Fig.~5 in \cite{Murray:2007p211} shows that $\chi({\rm Mg}^{+})$
lies between 0.1 and unity over broad range in gas density
in ULIRG outflows due to the very low ionization parameter in
dusty galaxies.

Using Cloudy version 13.00 \citep{Ferland:2013}, we calculated the photoionization 
equilibrium of a constant density gas slab illuminated by a starburst and examined
$\chi({\rm Mg}^+)$ as a function of $U$. Over the range  $8 \times 10^{-4} < U < 0.08$, 
a simple power law fit gives
\begin{eqnarray}
\chi({\rm Mg^+}) \approx n_{{\rm Mg^+}} / n_{{\rm Mg}} = \epsilon_{Mg+} U^{-\alpha},
\end{eqnarray}
with $\epsilon_{Mg+} = 4 \times 10^{-4}$ and $\alpha = 0.94$. Dropping the
assumption introduced in Eqn.~\ref{eqn:trick}, Equation~\ref{eqn:density}
is modified to 
\begin{eqnarray}
n_{H} \approx \left [
              \frac {n_{{\rm Mg^+}}}{ \eta({\rm Mg^+}) d({\rm Mg})  \epsilon_{Mg+}}
               \right ] ^{1/(1+\alpha)}
              \left [ \frac {Q}{4 \pi c r^2}
               \right ] ^{\alpha/(1+\alpha)}.
\end{eqnarray}
For these improved values of $\chi({\rm Mg}^+)$ and $\alpha$, Eqn.~\ref{eqn:nH} becomes
\begin{eqnarray}
n_{H} = 0.15 {\rm ~cm}^{-3} 
\left ( \frac{n_{{\rm Mg}^+}}{10^{-9} {\rm ~cm}^{-3} }  \right )^{0.52}
\left ( \frac{11.4 {\rm ~kpc}}{r_S} \right )^{0.96} \times \nonumber \\
\left ( \frac{6.3 \times 10^{-2}}{d({\rm Mg})} \right )^{0.52}
\left (\frac{3.8 \times 10^{-5}}{\eta({\rm Mg})} \right )^{0.52}
\left (\frac{Q}{10^{55} {\rm ~s}^{-1}} \right )^{0.48}.
\label{eqn:nH_cloudy} \end{eqnarray}
In the previous section, we argued that a sightline at impact 
parameter $b \approx 11.4$~kpc through the outflow includes
emission from $r_S \approx b$ up to  $r_S \approx 12 - 18$~kpc.
For this range of radii and $n_{Mg+} = 1.4 \times 10^{-9}$\cm3, 
our best estimate for the hydrogen density at the Sobolev radius
becomes
\begin{eqnarray}
n_{H} \approx 0.12 - 0.18 {\rm ~cm}^{-3}.
\end{eqnarray}

To facilitate comparison to TKRS~4389, we have adopted the same fiducial 
corrections for abundance and depletion as \cite{Rubin:2011p1660}. In contrast,
however, application of our argument for the ionization correction to TKRS~4389 will 
yield a significantly larger H density than the \cite{Rubin:2011p1660} estimate of
$n_H \approx 0.003$\cm3 because they assumed $\chi({\rm Mg}^+) \approx 1$. Using the
ionic density $n_{Mg+} = 2.6 \times 10^{-9}$\cm3 (from the previous section) in
Eqn.~\ref{eqn:nH}, we estimate $n_H = 0.28 - 0.67$\cm3 for $r_S = 20 - 8$~kpc. 
Over this same range of plausible values for $r_S$, our more accurate estimate
from Eqn.~\ref{eqn:nH_cloudy} is $n_H = 0.15 - 0.33$\cm3. 
We conclude that the gas density at the Sobolev points in the TKRS~4389 
and 32016857 outflows are very similar.

\subsection{Mass-loss Rate}

Using the {\rm H} gas density from the previous section and assuming 1.4 atomic 
mass units per hydrogen atom, the mass loss rate in the warm gas from Eqn.~\ref{eqn:conserve}
becomes
\begin{eqnarray}
\dot{M}(r) = 500 {\rm ~M_{\odot} ~yr}^{-1} 
\left (\frac{\Omega}{\pi} \right )
\left ( \frac{f_c}{1} \right )
\left ( \frac{r}{11.4 {\rm ~kpc}} \right )^2 \times \nonumber \\
\left (\frac{v}{230 {\rm ~km~s}^{-1}} \right ) 
\left ( \frac{n_H}{0.3 {\rm ~cm}^{-3}} \right ).~~
\label{eqn:mdot} \end{eqnarray}
We have used the observation that resonance lines are blueshifted 100\kms 
or more in 20\% of star-forming galaxies at $z \sim 1$ \citep{Martin:2012}, 
and the interpretation that the outflows must subtend roughly
$0.2 \times 4\pi = 0.8 \pi \approx \pi$ steradians on average. For
32016857, the mass flux in the (smooth) warm outflow is therefore approximately
330 - 500\msunyr, which is 4 - 6 times larger than the SFR of 80\msunyr\ 
(for the Chabrier IMF). For TKRS~4389, we estimate a mass flux of
35 to 40\msunyr.

We can use the empirical limits on the \mgII\ column density as a 
consistency check on the above argument. As can be seen in Figure~\ref{fig:mcmc}, 
resonant emission fills in the \mgII\ absorption troughs, so the
absorption equivalent width must be estimated from a fitted model.
Even the weaker transition of the \mgII\ doublet is optically thick 
in any reasonable fit to the 32016857 spectrum; and, in Figure~\ref{fig:mcmc},
it has an equivalent width of $W_{2803} = 2.04$ \AA. We can
use the linear relationship between $W_{2803}$ and column density,
valid  in the optically thin limit, to place a lower limit on the 
ionic column density of $N({\rm Mg}^+) > 1.0 \times 10^{14}$\col.

Substituting Eqn.~\ref{eqn:conserve} into Equation~\ref{eqn:N}
defines the relationship between this column density and the mass-loss rate.
For a constant velocity outflow, $v(r) = V_{Dop}$, it yields
\begin{eqnarray}
\dot{M} \approx \Omega \bar{m}_{ion} f_cs N_{ion} V_{Dop} R_0.
\end{eqnarray}
The mass-loss rate estimated from absorption lines therefore scales
linearly with the inner radius of the outflow, $R_0$, which physically represents a 
launch radius.  Using our mass-loss rate estimate from the scattered emission
and the lower limit on the column density, however, we obtain an upper limit
on the maximum launch radius
\begin{eqnarray}
R_0 <  2.0 {\rm ~kpc} 
\left ( \frac{\dot{M}}{500 \msunyr} \right )
\left ( \frac{\pi}{\Omega} \right ) 
\left ( \frac{7 \times 10^{-16} {\rm ~g}}{\bar{m}} \right ) \times \nonumber \\
\left ( \frac{10^{14.5} \col}{N_{Mg+}} \right )
\left ( \frac{230 \kms}{V_{Dop}}  \right ).~~~~~
\end{eqnarray}
Dynamical models of outflows relate the launch radius to 
the disk scaleheight \citep{MacLow:1989p141,Murray:2011p66},
suggesting values of $R_0$ ranging from $\approx 100 {\rm ~pc}$ for thin gas disks
to several kiloparsecs for dwarf galaxies and very turbulent systems.
We conclude that the strength of the saturated absorption lines, while in
no way providing a test of our large ionization correction, are fully 
consistent with the ionic density of ${\rm Mg}^+$ inferred from the scattered
emission.

Up to this point, we have tacitly assumed a smooth outflow with $f_c \approx 1$.
Higher resolution spectroscopy of outflows, however, indicates that saturated, 
low-ionization lines are not always black; and this partial covering of the
continuum light indicates the warm outflows do not always have unity covering
fraction \citep{Martin:2009}. From a physical standpoint, interstellar gas is 
likely entrained in a hotter wind to create the outflow. In the Appendix to
this paper, we outline simple physical arguments for the density and number
of warm clouds in outflows. We find that the clumps have higher $n_H$ than
the smooth wind but reduce the overall mass-loss rate. The case of clouds
confined by a hot wind produces higher gas densities in the outflow than do
unconfined clumps. Both models suggest that clumps may lower the inferred
mass-loss rate by up to a factor of 10. We conclude that the mass-loss
rates in warm outflows from 32016857 and TKRS~4389 are at least 30-50\msunyr
and 35-40\msunyr, respectively. In this multi-phase model of the wind, 
we find the mass flux in the warm phase alone is comparable to the SFR.

\section{Implications}

One of the major goals of feedback studies is to measure the rate at which
galactic winds remove gas from star-forming regions. 
Our empirical understanding of the relative mass flux originated with
the study of starburst galaxies nearby enough to directly detect their
winds in X-ray emission. The mass flux
in both the $> 10^7$~K \citep{Martin:2002,Strickland:2009p697}
and  $10^4$~K \citep{Lehnert:1996p651,Martin:1999} phases of the winds was shown to be
of order the SFR.  At higher redshifts where winds presumably had an
even larger impact on galaxy growth, the hot phase can rarely be directly
observed; but unique observations of the warm-hot plasma now indicate it may
contain 10 to 150 times more mass than the cold gas \citep{Tripp:2011p952}.

In this paper we demonstrated how the resonance emission scattered by the
warm gas in galactic winds can be used to measure the mass flux of low-ionization 
gas. Our discussion focused on the $z = 0.9392$ galaxy 32016857 because we detected 
\mgII\ emission that is significantly more extended than the nebular [\ion{O}{2}] 
emission and marginally more extended than the near-UV continuum emission.
This halo emission almost certainly results from photon scattering because
the Doppler shift of the \mgII\ emission is constant along the slit and
does not share the prominent velocity gradient seen [\ion{O}{2}] emission. The halo
gas appears to be an outflow based on the P-Cygni line profile of \mgII\ in 
the integrated spectrum and the blueshift of the \feII\ resonance absorption lines.
This outflow must be collimated (i.e., an opening angle $\Omega < 4\pi $) because (1) the \mgII\ emission
is only extended in one direction (to the east) along the slit and (2) the optical depths of the 
\mgII\ emission and absorption components differ, indicating departure from spherical symmetry.

Resolving the scattered emission improves our understanding of the mass-loss rate
in the low-ionization phase of the outflow. In the Sobolev approximation, the 
ionic density and the radial velocity gradient determine the scattering optical depth
along a sightline through the outflow. Based on the Doppler shift of the
low-ionization (\feII) absorption and the spatial extent of the \mgII\ emission
in the 32016857 spectrum, we estimate an average velocity gradient of 230 \kms\ 
over 12 to 18~kpc for the outflow. Assuming we have detected the surface 
of unity scattering optical depth, an ansatz consistent with the doublet ratio of 
the \mgII\ emission, we directly calculate the density of Mg$^+$ ions, 
$n(\mgII) \approx (0.89 - 1.4) \times 10^{-9}$\cm3, in the outflow
well beyond the star-forming region.

The implied mass flux depends on assumptions about metallicity, ionization fraction,
depletion, and opening angle. We adopted a typical opening angle derived from the fraction
of spectra of $z \sim 1$ star-forming galaxies showing blueshifted resonance absorption
\citep{Martin:2012} and a metallicity that ensures consistency with the mass -- metallicity
relation at similar redshift \citep{Zahid:2011p137}.  The depletion of Mg onto grains
is less constrained, but we argue that the likely values span only a factor of four
based on the well-measured properties of Galactic clouds \citep{Savage:1996p1461}. The
most uncertain parameter was the ionization correction, and we used
photoionization modeling to show that the relation between the ionization fraction and
the ionization parameter is nearly linear in the physical regime of interest. This
calculation strongly suggests that most of the Mg in the outflow is ${\rm Mg}^{+2}$ and
therefore not directly detected by our spectra.  Following these physically plausible 
arguments, the mass flux in a smooth outflow is approximately  330 - 500\msunyr\ in 32016857,
where the estimated uncertainty is a factor of four and the ${\rm SFR} \approx 80$\msunyr. 
We reached very similar conclusions for TKRS~4389 outflow using analogous arguments.

Since we expect winds to be multi-phase, we further argued that the clumpiness of 
the warm gas must be taken into account to accurately characterize the systematic uncertainty 
in the mass-loss rate. In an Appendix, we demonstrated that smaller clouds result in higher
densities and lower mass-loss rates (relative to the smooth outflow). The requirement that
the clouds survive out to at least the observed distance of the scattering halo, however,
places a physical limit on the minimum size of the clouds; and we show  that outflow
models with clouds reduce the inferred mass-loss rate by a factor of ten 
relative to the smooth wind model. Accounting for clouds lowers the inferred mass-loss 
rate in the warm outflow  but allows additional mass-loss in the hot phase. 
For 32016857 and TKRS~4389, we reach a reasonably robust conclusion that 
the outflow rates in the warm phase are at least comparable to the SFRs of the
galaxies and could be an order of magnitude higher in the absence of a hot phase.

The mass loading factor, defined as $\eta \equiv \dot{M} / SFR$, serves a critical
function in models of galaxy evolution. It scales inversely with the halo 
velocity dispersion, $\sigma$, for momentum-driven winds, but for energy-driven
winds the relationship steepens to $\eta \propto \sigma^{-2}$. This slope affects
how quickly the faint-end of the galaxy mass function flattens 
\citep{Creasey:2013p437,Puchwein:2013p2966} as well as the slope of the mass-metallicity 
relation \citep{Dave:2012p98,Shen:2012p50}. To fit the galaxy
mass function and mass -- metallicity relation, the mass flux associated 
with a given star formation rate needs to increase as galaxy mass decreases.
The typical normalization in numerical simulations sets the mass-loss rate to
roughly the SFR at some fiducial mass. The relative mass fraction at widely 
different temperatures in these winds is not yet a robust theoretical prediction. 
Simulations focusing on the winds rather than cosmological structure have 
just begun to include pressure from radiation and cosmic rays which analytic 
calculations suggest are important \citep{Murray:2005p569,Murray:2011p66,Breitschwerdt:1991,
Everett:2008p258,Socrates:2008}. At least in one recent simulation, \cite{Hopkins:2012p3522}
find that the warm, $2000 < T (K) < 10^5$, phase dominates the mass transport 
in outflows from SMC-like dwarf galaxies and remains comparable to the mass flux 
in the hot wind in a model chosen to match the properties of a high-redshift starburst.

At this time, TKRS~4389 \citep{Rubin:2011p1660} and 32016857 remain the only clear 
examples of \mgII\ scattered by outflows.
Among 145 LRIS spectra of $0.4 < z < 1.4$ galaxies, including 22 with very prominent \mgII\ 
emission, we found only three examples of spatially resolved \mgII\ emission. Unlike 32016857,
the line emission resolved around 32010773 and 12019973 is not necessarily scattered;
it could come directly from \ion{H}{2} regions as it is no more extended than the
nebular [\ion{O}{2}] emission and not clearly distinct kinematically. 
We suggest, in spite of the large number of null detections,  that diffuse \mgII\ emitting 
halos may be a common property of star-forming galaxies at intermediate redshifts. Approximately
15\% of our sample have prominent \mgII\ emission \citep{Kornei:2013}. A larger fraction of
the spectra show blueshifted resonance absorption lines indicative of outflowing gas \citep{Martin:2012}.
The resonance emission scattered by these outflows could be attenuated by the LRIS slit or
destroyed by dust absorption, a scenario supported by the observed tendency for \mgII\ 
emitters to have bluer than average $U-B$ color \citep{Martin:2012,Kornei:2013}.
Based on the properties of TKRS~4389 and 32016857 -- namely their high sSFRs and unusually 
high B-band luminosities for their $U-B$ colors, we suggest that the scattered emission
is brighter around these galaxies than it is near  more typical \mgII\ emitters. 

While much of their halo emission may simply lie below current detection limits, intermediate 
redshift galaxies may prove an ideal environment for pinning down mass-loss rates. 
Outflows were more common at $z \sim 1$ than today due to the higher cosmic SFR \citep{Hopkins:2006p142} 
and smaller galaxy sizes \citep{Simard:1999p563, Lilly:1998p75}. Most of the 
star-forming galaxies have developed well-defined disks by this time. The orientation of 
these galactic disks relative to halo sightlines probed by background quasars (or galaxies)
provides a valuable means to distinguish intervening absorption created by outflows
from other sources such as infalling gas \citep{Bordoloi:2011p1640}. The relative metallicity 
of the galactic disk and halo clouds have also been used to identify accreting gas
\citep{Ribaudo:2011}. Scattered emission from 
an outflow becomes easier to detect as the physical area subtended by a fixed spectroscopic
aperture increases, and the angular diameter distance increases with redshift out to $z \approx 1.6$. 
Since cosmic expansion reduces the observed surface brightness emission as $(1 + z)^{-4}$,
the brightest scattered halo emission may be found at redshifts  somewhat less than the
maximum angular diameter distance.
In the future, mapping the \mgII\ line emission scattered by circumgalactic gas 
should, in principle, provide a means to directly measure $\eta$ in galaxies over 
a broad mass range. The results from the first two detections of scattered \mgII\
emission strongly suggest very substantial mass-loss in the outflows from intermediate
redshift galaxies.

\acknowledgements
We thank the the DEEP2 and AEGIS teams for providing ancillary data on galaxy properties, 
Mike Shull for stimulating discussions, and the referee for thoughtful comments on the manuscript. 
This research was supported by the National Science Foundation under AST-1109288 (CLM),
the David \& Lucile Packard Foundation (AES and CLM), the Alfred P. Sloan Foundation 
and NSF CAREER award AST-1055081 (ALC), 
a Dissertation Year Fellowship at UCLA (KAK), and an NSF Graduate Research Fellowship Program (AP).
The research was partially carried out at the Aspen Center for Physics which is
supported by the National Science Foundation under Grant No. NSF PHY05-51164.
We also wish to recognize and acknowledge the highly significant
cultural role that the summit of Mauna Kea
has always had within the indigenous Hawaiian community. It is a
privilege to be given the opportunity to conduct observations from
this mountain.

{\it Facilities:} \facility{Keck}

\begin{deluxetable}{lc}
\tablecaption{Galaxy Properties}
\tabletypesize{\scriptsize}
\tablewidth{0pt}
\tablehead{
\colhead{Property} &
\colhead{32016857} 
}
\startdata
RA                                   & 23:29:24.954 \\
DEC                                  & +00:07:05.85 \\
Redshift\tablenotemark{a}            & 0.939196  \\
$B$ (AB mag)                                  & 22.49     \\
$M_B - 5 \log h_{70}$ (AB mag)\tablenotemark{b}    & -21.59    \\
$U-B$\tablenotemark{b}               & 0.44      \\
$\log M_* / \msun$\tablenotemark{c}  & 9.82      \\
$V_1$ (\kms)\tablenotemark{a}        & $-91 \pm 15$ \\
$V_{Dop}$ (\kms)\tablenotemark{a}        & $-227 \pm 82$ \\ 
$\log [N_{Dop}(Fe^+) C_f]$ (\col)\tablenotemark{a}   & 14.41 - 15.53 \\
SFR (\msunyr)\tablenotemark{d}                      & 78 \\
sSFR (yr$^{-1}$)                      & $1.2 \times 10^{-8}$ 
\enddata
\tablenotetext{a}{From \cite{Martin:2012}.}
\tablenotetext{b}{From \cite{Willmer:2006p1661}.}
\tablenotetext{c}{From \cite{Bundy:2006p1663}.}
\tablenotetext{d}{After scaling the flux of the red LRIS spectrum
to match the DEEP2 magnitude of $I_{AB} = 21.70$, we measured an
integrated [\ion{O}{2}] line flux of $6.25 \times 10^{-16}$\flux,
corresponding to an uncorrected (for extinction) line luminosity
of $2.79 \times 10^{42}$~erg~s$^{-1}$. To estimate the SFR, we
use the average ratio of [\ion{O}{2}] to \Ha\ luminosity (0.40) 
measured by \cite{Moustakas:2006p775} among nearby galaxies
with $M_B \approx -21.6$. By adopting the mean ratio for
galaxies with luminosities similar to 32016857, this calibration
takes the strong correlation between line ratio and reddening
into account. Finally, dividing the SFR obtained from 
the \citep{Kennicutt:1998araa} \Ha\ calibration by 1.78, we
obtain a SFR of 78\msunyr\ for a Chabrier IMF from 0.1 to 100\msun.
}
\label{tab:galaxy}  
\end{deluxetable}
\normalsize

\begin{deluxetable}{llcc}
\tablecaption{Spectral Line Measurements}
\tabletypesize{\scriptsize}
\tablewidth{0pt}
\tablehead{
\colhead{Transition} &
\colhead{$\lambda_0$}  &
\colhead{$W_r(\lambda)$}  
\\
\colhead{} &
\colhead{(\AA)}  &
\colhead{(\AA)}  
}
\startdata
\feII$^*$  & 2365.56     & $-0.47 \pm 0.34$           &    \\         
\feII$^*$  & 2381.49     & \nodata                    & \\
\feII$^*$  & 2396.26     & $-0.78 \pm 0.32 $           &    \\         
\feII$^*$  & 2612.65     & $0.31 \pm 0.33$           &    \\         
\feII$^*$  & 2626.45     & $-1.41 \pm 0.32$           &    \\         
\feII$^*$  & 2632.77     & $-0.32 \pm 0.24$           &    \\         
\hline
\feII      & 2260.21     & $< 0.6$                     & \\ 
\feII      & 2344.21     & $ 1.41 \pm 0.22$           &    \\         
\feII      & 2374.46     & $ 0.88 \pm 0.19 $           &    \\         
\feII      & 2382.77     & $ 0.99 \pm 0.18 $           &    \\         
\feII      & 2586.65     & $ 1.84 \pm 0.22 $           &    \\         
\feII      & 2600.17     & $ 2.07 \pm 0.24 $           &    \\                      
\mgII      & 2796.35     & $ 0.34 \pm 0.39$          &    \\                      
\mgII      &   ''        & $-2.13 \pm 0.29$\tablenotemark{b}          &    \\                      
\mgII      & 2803.53     & $ 0.85 \pm 0.22$          &    \\                      
\mgII      &   ''        & $-1.16 \pm 0.27$\tablenotemark{b}          &    \\                      
\hline
CIII]                                  & 1908.73                    & $-1.57 \pm 0.47$           &    \\
CII]                                   & 2326\tablenotemark{a}      & $-2.06 \pm 0.37$           &    \\         
\[[\ion{Ne}{5}]                               & 3426.98                    & $< 0.84$           &    \\                      
\[[\ion{O}{2}]                               & 3727.09, 3729.88           & $-128.2 \pm 3.1  $           &    \\                       
\[[\ion{Ne}{3}], \ion{He}{1}                       &  3869.85, 3868.63          & $-14.16 \pm 0.95 $           &    \\                      
H8, \ion{He}{1}                        &  3890.15, 3889.75          & $-6.83 \pm 0.63 $           &    \\                      
\ion{H}{1}, [\ion{Ne}{3}], \ion{He}{1} &  3971.19, 3968.58, 3965.85 & $-8.73 \pm 1.03 $           &    \\                      
H$\delta$                              &  4102.90                   & $-9.55 \pm 0.70 $           &    \\                      
H$\gamma$                              &  4341.69                   & $-16.68 \pm 1.36  $              
\enddata
\tablenotetext{a}{Blend of $\lambda 2324.22$, $\lambda 2325.41$, $\lambda 2327.65$, $\lambda 2328.84$}
\tablenotetext{b}{In the integrated spectrum, the observed flux in \mgII\ emission
is $F(\lambda 2796) \approx 1.5 \times 10^{-17}$\flux\ and is $F(\lambda 2803) \approx 1.0 \times 10^{-17}$\flux\
to roughly 20\% accuracy.}
\label{tab:lines}  
\end{deluxetable}
\normalsize

\clearpage

\appendix

\section{A. Mass-Loss Rate for a Clumpy Wind}

The scattering optical depth of the $\lambda 2796$ line is
\be \label{eqn: optical depth}
\taumg = n_{\rm Mg II}\sigmg (\Delta v)\, l.
\ee 
The cross-section of the $\lambda2796$ transition, with units
of $\,{\rm cm}^2$, is
\be 
\sigmg (\Delta v) ={\pi e^2\over m_e c}f{\lambda_0\over \Delta v},
\ee 
where $\lambda_0$ is the wavelength of the transition, $e$ and $m_e$
are the charge and mass of an electron, and $f=0.62$ is the oscillator
strength of the transition. The quantity $\Delta v$ is the velocity
range over which the ions are distributed; for example, where the gas
exhibits only thermal motions, $\Delta v=\vth$, so that the velocity
spread is simply the thermal velocity spread. The velocity spread may
be much larger than $\vth$ if the gas exhibits bulk motions.

In fact the expression (\ref{eqn: optical depth}) for the optical
depth depends on two unknown quantities, a characteristic length scale
$l$ associated with the outflow, and $\Delta v$. We scale $\Delta v$
to the ion thermal velocity $\vth \approx  2 (T/10^4\K)^{1/2}$ \kms 
of a ${\rm Mg}^+$ ion, and introduce
\be 
\sigth={\pi e^2\over m_e  c}f{\lambda_0\over\vth}\approx2.6\times10^{-12}
\left({T\over10^4\K}\right)^{-1/2}
\,{\rm cm}^2.
\ee 
Using these, eqn. (\ref{eqn: optical depth}) becomes
\be 
\taumg = n_{\rm Mg II}\,\sigth {\vth\over\Delta v}\, l = 
n_{\rm Mg II}\,\sigth\, l_{\rm eff},
\ee 
where we have introduced the effective scattering length
\be 
l_{\rm eff}\equiv {\vth\over \Delta v}l.
\ee 

Specifying a physical model for the outflowing gas determines $l_{\rm
  eff}$. We will scale to a smooth wind model, in which the effective
scattering length is the Sobolev length
\be 
l_{\rm Sob}\equiv{\vth\over dv/dr},
\ee 
where $dv/dr \approx v(r) / r_S$ is the velocity gradient in the wind at the point where $\taumg=1$.

For our galaxy, taking the largest radius at impact parameter $b$ where scattered
emission is detected,
\be 
\lsob\approx \left({\vth\over v}\right)\approx 90\pc
\left ( \frac{\vth}{2 {\rm ~km~s}^{-1}} \right )
\left (  \frac{280 {\rm ~km~s}^{-1}}{v} \right )
\left ( \frac{r_S}{12 {\rm ~kpc}} \right ).
\ee 

\subsection{A.1. Physical wind models}
We consider three models for the outflow, a smooth $T\approx10^4\K$
wind, an outflow consisting of unconfined $T\approx10^4\K$ clouds or
clumps, and a two-phase hot ($T_h\approx10^8\K$ and smooth) together
with a cold ($T\approx10^4\K$) and clumpy outflow, in which the cold
clouds are confined by thermal pressure of the hot gas. The hot gas in
the last model is naturally provided by supernovae exploding in the
galaxy, at a rate proportional to the star formation rate (about one
supernova per century for a star formation rate of one solar mass per
year).

\subsubsection{A.1.1. The smooth wind model}
In a smooth wind the characteristic velocity spread is
$\Delta v=\vth$ and the characteristic length $l$ is 
the Sobolev length
\be 
l_{\rm eff} = l_{\rm Sob}\equiv {v_{th}\over dv/dr}.
\ee 
The logarithmic derivative of the wind velocity
\be 
\Gamma(r)\equiv {r\over v}{dv\over dr}
\ee 
is of order unity in the wind acceleration region, so 
\be 
l_{\rm Sob}=\Gamma^{-1}\left({\vth\over v}\right)r<< r.
\ee 

\subsubsection{A.1.2. Unconfined clouds}

Our second model assumes that the outflow consists of a large number of
unconfined clouds. If the clouds are unconfined, e.g., if their gas pressure
exceeds that of any hot outflow, then their size along the direction
of acceleration ${\bf g}$ is
\be 
l={c_s^2\over g}\approx\left({c_s\over v_c}\right)^2 r\approx100\pc,
\ee 
where $c_s\approx10\kms$ is the sound speed of the $T\approx10^4\K$
gas in the clouds, and $v_c$ is the circular velocity of the host
galaxy, which is of order the outflow velocity. The effective size of
the cloud is
\be 
l_{eff}=
\left({c_s\over \Delta v}\right)
\left({c_s\over v} \right)
\left({\vth\over v}\right) r,
\ee 
where we approximate $v_c \approx v$, the outflow velocity.

The velocity spread $\Delta v\approx \vth$ if there are no bulk flows
in the clouds, in which case $l_{eff}=l_{\rm Sob}$. In the more likely case that
there are transverse bulk flows (expansions, in particular), then
$\Delta v$, could be of order $c_s$. In that case,
\be 
l_{eff}\approx {c_s\over v}\lsob\approx0.1\lsob.
\ee 

\subsubsection{A.1.3. Confined clouds}
Our third model assumes that the cold gas consists of clouds confined
by hot gas. With a star formation rate of $\sim80M_\odot\yr^{-1}$, and
an expected supernova rate of nearly one per year, it is likely that
the galaxy possesses a hot ($T_h\sim10^8\K$) gas component. Since the
sound speed of this hot gas exceeds the circular velocity of the
galaxy by a large amount, it will flow out with a bulk velocity of
a few times its sound speed, $c_h\approx1000\kms$. This gas will
act to pressurize the cold gas clouds, to accelerate them, and to
disrupt them.

A priori, the cold clouds could have any size $l_{cl}$ and a range of
internal velocities $\Delta v$. However, we argue that there is a well
defined lower limit to $l_{cl}$, as follows.

Both simulations \citep{Klein:1994p213} and experiments
\citep{Hansen:2007p379} demonstrate that cold clouds embedded in
such hot outflows are destroyed on a few tens of cloud crushing
times. The cloud crushing time is defined as
\citep{Klein:1994p213}
\be 
\tau_{cc}\equiv\left({n_c\over n_h}\right)^{1/2}
{l_{cl}\over v_s},
\ee 
where $n_c=n_H^{\tau=1}$ and $n_h$ are the number densities of the cold and hot
gas, and $v_s\approx v_h$ is the velocity of the shock in the hot
flow.

The clouds must live long enough to reach $R_{\tau=1}\approx11.4\kpc$,
which allows us to set a lower limit to $l_{cl}$:
\be 
30\tau_{cc}\gtrsim 
\left(R_{\taumg=1}\over v \right),
\ee 
leading to a cloud length
\be 
l_{cl}\gtrsim 
\left({v_h\over30 v}\right)^{4/3}
\left({n_h\over n_H^{\tau=1}({\rm Sob})}\right)^{2/3}
\left({r\over\lsob}\right)^{4/3}
\left({\vth\over\Delta v}\right)^{1/3}
\lsob.
\ee 
We have scaled the clouds survival time to $30$ cloud crushing times.
Scaling to a hot wind with a mass loss rate similar to the star
formation rate, and assuming that $\Delta v\sim\vl$,
\be 
l_{cl}\approx170
\left({R\over 11.4\kpc}\right)
\left({n_h\over 2\times10^{-3}\,{\rm cm}^{-3}}\right)^{1/2}
\pc.
\ee 

As with the unconfined cloud model, this length is similar to the
Sobolev length associated with a smooth wind. The characteristic
velocity spread $\Delta v\sim \vl$; it cannot be larger than the
observed linewidth. Then 
\be 
l_{eff}\equiv\left({\vth\over v}\right) l_{cl}
\approx 1.7\pc. 
\ee 

If the gas being removed from the cloud is
over-ionized (so that all the MgII ions are removed) before it is
accelerated, $\Delta v$ could be as small as $\vth$, but as we will see
this would lead to a large mass loss rate in the cold component.

\subsection{A.2. Mass-loss rates}

For a smooth outflow, the continuity equation 
\be \label{eqn:continuity2} 
\dot M_w = \Omega f_c
\bar m r^2n_H v
\ee 
gives Eqn.~\ref{eqn:nH}.
Note that $\eta_{Mg}$, $d_{Mg}$, and $\epmg$ all enter to the one half
power, so that the estimate for the mass loss rate is only moderately
sensitive to these quantities. Observations indicate, however, that the 
warm outflows are clumpy \citep{Martin:2009,Bordoloi:2012p3774}.
Taking $f_c \approx 1$, where $f_c$ describes the fraction of
the outflow cone covered by clouds as seen from the galaxy,
places an upper limit to the mass loss rate in warm ($10^4\K$) gas
for a fixed density, $\bar{m} n_H$.

For illustration, we consider a bi-conical outflow of half-opening angle $\theta_{bc}$ such
that the outflow subtends a solid angle $\Omega = \Omega_{bc} \equiv 2 \pi \int_0^{\theta_{bc}} \sin
\theta d \theta$. In light of the limited observational constraints on the orientation
of the outflows in 32016857 and TKRS~4389, the symmetry axis of the
cone can be taken perpendicular to the observer's sightline to illustrate the impact of
clumpiness on the density and mass-loss rate.

\subsubsection{A.2.1. Clumpy outflows}
The mass of a single cloud is
\be 
m_{cl}={4\pi\over3}l_{cl}^3\bar m n_{H,cl}.
\ee 
Over the velocity width of the emission line, $\Delta v_{el}$, clouds at the projected 
radius where $\taumg (b) \approx1$ have to reflect a large fraction of the continuum. 
To find the number of clouds at this impact parameter which have projected velocities
consistent with the linewidth, consider an annulus of width $\Delta b$ on the sky. Over
the velocity width of a single cloud, $\Delta v_{cl}$, the clouds must cover a 
projected area $2 \theta b \Delta b$. In addition, the clouds have 
to cover the range of velocity seen in the emission line. Together these two constraints
give a lower limit to the number of clumps or clouds given by
\be 
N_{cl}=2  f_v
\left ( \frac{\theta_{bc}}{\pi} \right )
\left({\Delta v_{el} \over \Delta v_{cl}}\right)
\left({b\over l_{cl}}\right)
\left({\Delta b\over l_{cl}}\right),
\ee 
where 
$f_v$ is
the fraction of the linewidth covered by clouds.
The total mass in clouds is $M_{cl}=N_{cl}m_{cl}$, or
\be 
M_{cl} = 2 
\left ( \frac{\theta_{bc}}{\pi} \right )
f_v
\left({\Delta v_{el}\over \Delta v_{cl}}\right)
\left({b\over l_{cl}}\right)
\left({\Delta b\over l_{cl}}\right)
\cdot
{4\pi\over 3}l_{cl}^3\bar m n_{H,cl}.
\ee 

The mass loss rate is calculated using the crossing time, 
\be 
\tau_{\rm cross}= {\Delta b\over v_x(b)} = {\Delta b\over v \sin \theta} =
{\Delta b\over v} \cdot {r\over b}.
\ee 
We find
\be \label{eq: clumpy mass loss}
\dot M_{cl} = \bar m r^2 v \cdot 4\theta_{bc} f_v n_{\rm H,cl}
\cdot
{2\over3} \cdot
\left({l_{cl}\over r}\right)
\left({b\over r}\right)^2
\left({\Delta v_{el}\over \Delta v_{cl}}\right).
\ee 
Using equation (\ref{eqn:continuity2}), we scale the mass-loss rate in the
clumpy outflow to the that of the smooth outflow,
\begin{eqnarray}
\dot{M}_{cl} =
\left ( \frac{4\theta_{bc}}{\Omega_{bc}}  \right )
\left ( \frac{f_v}{f_c} \right ) \cdot 
\left ( \frac{n_{H,cl}}{n_H} \right )
\cdot
\left ( \frac{2}{3} \right )
\left ( \frac{l_c}{r}\right )
\left ( \frac{b}{r}\right )^2
\left ( \frac{\Delta v_{el}}{\Delta v_{cl}} \right )
\dot{M}_w. 
\label {eqn:mcl_mw}\end{eqnarray}

It is convenient to scale the cloud density to a smooth wind:
\be \label{eqn: density scaling}
n_{H,cl} = n_H({\rm Sob})
\left({\lsob\over l_{eff}}\right)^{1/2},
\ee 
where
\be 
n_H({\rm Sob})\equiv
\left[
{n_\gamma\nmgii^{\tau=1}({\rm Sob})\over \eta_{Mg}d_{Mg}\epmg}
\right]^{1/2}. 
\ee 
For a clumpy outflow, the effective length scale
for absorption over the linewidth contributed by a single clump is 
simply related to the properties of a cloud,  
\be 
l_{eff} \approx 
\left ( \frac{v_{th}}{\Delta v} \right ) l_{cl}. 
\ee 
The density of a 
clump or cloud in the wind will exceed that of the smooth wind
by a factor
\be \label{eqn:density_ratio}
n_{H,cl} = n_H 
\left({\lsob\over l_{cl}}\right)^{1/2}
\left({\Delta v_{cl} \over v_{th}}\right)^{1/2}.
\label{eqn:density_contrast}\ee 

\subsubsection{A.2.2. Comparison of Clumpy and Smooth Outflows}

Using equations (\ref{eqn:mcl_mw}) and (\ref{eqn:density_contrast}),
\be \label{eqn: clumpy scaling}
\dot M_{cl} =
{2\over 3}
\left ( \frac{4\theta_{bc}}{\Omega_{bc}} \right )
\left ( \frac{f_v}{f_c} \right )
\left({l_{Sob} \over r}\right)
\left({l_{cl} \over \lsob}\right)^{1/2}
\left({b \over r}\right)^2
\cdot
\left({\Delta v_{el} \over \Delta v_{cl}}\right)^{1/2}
\left({\Delta v_{el} \over \Delta v_{th}}\right)^{1/2}
\dot M_w.
\ee 
The square-root dependence of the mass loss rate on the cloud size $l_{cl}$
indicates this estimate is fairly robust. 
To illustrate how the clouds affect the mass-loss estimate, we
take $r = (1.0 - 1.5) b \approx b$, $4\theta_{bc} / \Omega_{bc} \approx 1$,
and $f_v / f_c \approx 1$.

For an unconfined clumpy outflow, with $l_{cl}\approx0.1\lsob$ and
$\Delta v\approx c_s$, the mass loss rate
\be 
\dot M_{cl} \approx 0.10
\left({\Delta v_{el} \over 280}\right)
\left({10 \over \Delta v_{cl}}\right)^{1/2}
\left({2 \over \Delta v_{th}}\right)^{1/2}
\left({l_{Sob} \over 90}\right)^{1/2}
\left({l_{cl} \over 9}\right)^{1/2}
\left({12 \over r}\right)
\dot M_w
\ee 
is about one tenth the smooth wind rate. 
The mass-loss is also lower when we assume hot gas confines the clouds.
For a confined clumpy wind, with $l_{cl}\approx\lsob$ but
$\Delta v\approx\vl$, the mass loss rate is 
\be 
\dot M_{cl} \approx 0.06
\left({\Delta v_{el} \over 280}\right)
\left({280 \over \Delta v_{cl}}\right)^{1/2}
\left({2 \over \Delta v_{th}}\right)^{1/2}
\left({l_{Sob} \over 90}\right)^{1/2}
\left({l_{cl} \over 90}\right)^{1/2}
\left({12 \over r}\right)
\dot M_w.
\ee 
Allowing for very small cloud sizes $l_{cl}$, as might be the case in a hot-gas 
confined clumpy outflow, indicates that the mass loss rate can up to a factor of 
approximately 10 times lower than that estimated under the smooth outflow assumption.

Whether small clouds can survive the acceleration process remains open
to some debate. The fiducial values of 9 and 90~pc used here lie at
the low end of the size ranges recently estimated from photoionization
modeling of circumgalactic gas clouds in both nearby galaxies
\citep{Stocke:2013p148} and intermediate redshift absorption-line systems
\citep{Meiring:2013p49}. These calculations offer a physical argument
nonetheless for the maximum amount by which the smooth outflow model
may overestimate the mass-loss rate. They suggest that the mass-loss
rate in the warm outflow could be just 10\% of the smooth wind mass-loss 
rate. If such clouds are present in the wind, then the hot wind will
of course contribute additional mass flux.

\bibliographystyle{apj}
\bibliography{Papers_winds}

\end{document}